\documentclass[10pt,letterpaper]{article}
\usepackage[top=0.85in,left=1in,right=0.8in,footskip=0.75in]{geometry}

\usepackage{amsmath,amssymb}

\usepackage{changepage}

\usepackage{textcomp,marvosym}

\usepackage{cite}

\usepackage{booktabs}

\usepackage{nameref,hyperref}

\usepackage{fancyvrb}

\usepackage{longtable}

\usepackage{float}

\usepackage[nopatch=eqnum]{microtype}
\DisableLigatures[f]{encoding = *, family = * }

\usepackage[table]{xcolor}

\usepackage{array}

\hypersetup{colorlinks=true, citecolor=blue, linkcolor=blue, urlcolor=blue}

\newcolumntype{+}{!{\vrule width 2pt}}

\newlength\savedwidth

\raggedright
\usepackage[aboveskip=8pt,labelfont=bf,labelsep=period,justification=raggedright,singlelinecheck=off]{caption}

\makeatletter
\renewcommand{\@biblabel}[1]{\quad#1.}
\makeatother

\usepackage{lastpage,fancyhdr,graphicx}
\usepackage{epstopdf}
\fancyheadoffset[L]{2.25in}
\fancyfootoffset[L]{2.25in}
\begin{document}
\vspace*{0.2in}

\begin{flushleft}
{\Large
\textbf\newline{PathwayDesigner: an interactive SBML/KEGG/Reactome/Escher compatible editor for biochemical network diagrams}
}
\newline
\
Herbert M. Sauro\textsuperscript{1,2}

\bigskip
\textbf{1} Department of Bioengineering, University of Washington, Seattle, 98195-5061, WA, USA

\textbf{2} eScience Institute, University of Washington, Seattle, 98195-5061, WA, USA

\bigskip

Corresponding author: hsauro@uw.edu

\end{flushleft}

\section*{Abstract}
PathwayDesigner is an interactive editor for biochemical network diagrams. A network
can be drawn directly on the screen, or described as an Antimony model, which
PathwayDesigner then lays out and draws. PathwayDesigner reads and writes SBML using the Layout and Render packages, writing
Render in a form that other tools can interpret correctly. PathwayDesigner can read and write a diagram's layout and appearance using either Antimony's own layout statements or in a new, more concise format that we propose. PathwayDesigner can also open Escher maps, KEGG maps and
Reactome's SBML-based maps. Other components of the model, such as events, rules, functions, and
units, are kept and can be edited within the application. Species and reactions can be annotated with references to any database, and PathwayDesigner can find and add annotation references automatically by searching the BiGG, ChEBI, PubChem, and Rhea databases with a component's Id or Display Name, or contextually with a reaction's reactants and products. The network's stoichiometry matrix, conservation laws, and flux null space are computed and visualized on the diagram. The visual style of a diagram can be set with named styles or themes, saved and reused across models. One built-in theme follows the Escher style. The software is written in pure Rust
as one code base for Windows, macOS, Linux, and the web browser, where it runs as
WebAssembly. The web version is hosted on a GitHub page so that a user doesn't need to manually install any code. All 1,062 curated BioModels models and the 1,823
semantic cases of the SBML Test Suite were read and written back without change. Diagrams were exchanged in both directions with COPASI and SBMLNetwork. PathwayDesigner is released as open source under
the MIT licence, at \url{https://github.com/sys-bio/PathwayDesigner-rs}, and runs in
the browser at \url{https://sys-bio.github.io/PathwayDesigner-rs/}.

\section*{Author summary}
This articles describes a new software application that can be used to edit the network diagrams of biochemical pathways found in living cells. The application supports a number of community standards such as SBML and Escher. This means that diagrams developed with this tool can be reused by other applications that wish to display the biochemical networks. The SBML standard also stores a computational and simulatable model of the biochemical system. This means that the computer model and a visual representation of the model reside in one unified file.

\section*{Introduction}

Biochemical network diagrams are used frequently for representing, communicating, and analyzing the complex relationships among molecular species and biochemical reactions. However, developing an effective network editor requires balancing intuitive graphical manipulation with the preservation of biochemical semantics, mathematical model structure, and interoperability with established modeling tools and file formats.

One of the first network editors was JDesigner~\cite{sauro2003next} which enabled the graphical construction of biochemical reaction networks. Since then more specialized process-diagram editors have merged such as CellDesigner~\cite{funahashi2008celldesigner}, which introduced structured representations of biochemical systems, and more recently to web-based tools such as Newt~\cite{balci2021newt}, which support standards-based biological diagram editing and exchange. Despite these advances, different editors continue to reflect distinct priorities in model construction, diagram representation, and software interoperability. Building on the JDesigner lineage, we present PathwayDesigner, a new biochemical network editor designed to address these challenges by combining direct graphical manipulation of network components with support for structured biochemical models and interoperable representations. The editor aims to provide a practical environment for constructing, annotating and modifying biochemical networks while maintaining the connection between their visual representation and underlying model structure.

The Systems Biology Markup Language (SBML)~\cite{hucka2003systems} provides a machine-readable format for representing biochemical models, including species, reactions, and kinetic laws, enabling models to be exchanged among compatible simulation and analysis tools. However, the core SBML specification does not fully define how a model should be arranged visually. The SBML Layout package~\cite{gauges2015systems}, addresses this limitation by representing graphical objects and their spatial relationships, including the positions and dimensions of species and reaction glyphs and the geometry of connecting curves. The complementary SBML Render package~\cite{bergmann2018sbml} defines graphical styles and visual properties, such as colors, line widths, and fills, allowing diagrams to retain more of their appearance across applications. Together, these standards support the exchange of both model structure and graphical information, although complete preservation of diagram appearance and editing behavior still depends on the extent to which individual tools implement the relevant specifications, especially the render package. These standards provide an important foundation for editors that seek to combine flexible graphical manipulation with reproducible, portable representations of biochemical networks. The editor described in this article is one of the first interactive editors to provide combined layout and render editing capabilities for SBML.

Other standards include SBGN~\cite{novere2009systems} but this is largely a visualization standard without a modeling component and will not be considered further.

\section*{Design and implementation}

\subsection*{Design objectives}

The main objective of this project was to develop a modern cross-platform and practical platform for editing SBML based network diagrams that use the layout and render standards. In addition, the project offered an opportunity to leverage modern AI assistance to build a pathway editor that one developer could maintain and extend, an aspect we already experienced with the Iridium platform~\cite{iridium2026}.

The author has written previous network editors such as JDesigner~\cite{sauro2003next}, however, JDesigner runs only on Windows and is no longer supported and never supported the layout or render standard. A more recent prototype was written but never formally released called PathwayDesigner~\cite{PathwayDesignerPrototype}.  A key limitation of the prototype was that it was limited to running on Windows and the Mac and both Linux and Web based solutions were unavailable. It was therefore decided to reexamine the development workflow as well as the editor's functionality. As a result the following four objectives were established:

\begin{enumerate}
\item \textbf{Exchange.} The editor will read and write the following formats: SBML with layout and render~\cite{gauges2015layout} and
Antimony~\cite{heydarabadipour2026antimony3}. In addition there will be read capabilities for Escher
maps~\cite{king2015escher} and Reactome~\cite{milacic2024reactome} exported SBML. The application will have limited support for KEGG maps, because KEGG maps don't have reaction curve information and access is difficult. As with many editing tools, the kinetic model (rate laws, parameters, initial values, rules and events) is kept with the drawing, so a file written by PathwayDesigner can be
simulated by other tools. We will also define (Appendix A) a subset of the SBML render standard that most tools currently support.
\item \textbf{Building and distributing.}  Another key objective is to also create a single code base that can be run on all four platforms, Windows, Mac, Linux and Web.  For this Rust was chosen as the development language based on its modern tool chain. Applications written in
Rust~\cite{rustlang,Sauro2025Rust}, can be built for the browser as WebAssembly~\cite{haas2017webassembly}
and for the desktop.  By combining this with GitHub Actions and GitHub Pages, which are free for open-source projects, there is seamless and automatic compilation across all platforms with little work required by the developer.  The advantage of the Web version is that users have nothing to install. One could argue that Rust is a difficult language to master (which it is~\cite{Sauro2025Rust}) which negates its utility, however with modern AI this is no longer an issue.
\item \textbf{A simple notation.} While some platforms such as CellDesigner~\cite{funahashi2008celldesigner} or Newt~\cite{balci2021newt} offer complex interfaces for creating diagrams~\cite{novere2009systems}, for PathwayDesigner is was decided to make it as simple as possible for a user to use. Thus species are drawn from a limited range of shapes and reactions are arcs or straight lines. To minimize friction, drawing of reactions requires minimal mouse action. For example, a reaction A~$\rightarrow$~B can be drawn by placing two species and clicking the reactant and then the product. This makes drawing diagrams fast and easy.
\item\textbf{Annotation.} Many communities rely on models begin annotated to unambiguously identify the various components in a model. It was decided to add reasonably comprehensive support for annotating and searching for annotations in PathwayDesigner so that users can examine annotations, add annotations,  and notes to a model, species, reactions and compartments.
\end{enumerate}

Because the model and its drawing are one document, three rules apply throughout. The
drawing does not contradict the model: every species and reaction that is drawn is
part of the model, and every species a rate law depends on is shown as connected to
its reaction. Appearance (colours, shapes, line widths, fonts and themes) belongs
to the drawing alone, so it can be changed without changing the model.

PathwayDesigner does not simulate models. Simulation is left to existing tools such as
COPASI~\cite{hoops2006copasi}, Tellurium~\cite{somogyi2015libroadrunner} or Iridium\cite{iridium2026}. We did however add the ability to perform some limited stoichiometric analysis which can exploit the visual representation.

\subsection*{Representation of a network}

\noindent{\bf Species and aliases.} A species is drawn as a node that includes drawing a box, circle,
ellipse or hexagon with its Id or display name inside or beside it. Each species has one primary
node, which carries its biochemical data (initial value, compartment and whether it is
a boundary species), and any number of aliases. An alias is a second drawing of the
same species elsewhere on the canvas. It has no data of its own and is never written
out as a separate species. Aliases let a heavily used species, such as ATP, be drawn
next to each reaction that uses it instead of at the center of a web of reaction lines.
Commands are provided to automatically split a species into aliases, one for each of its reactions. In addition, aliases can be merged back into their species. Null nodes mark an empty side of a reaction (a source or a sink) and are not species of the model.

\noindent{\bf Reactions.} A reaction is a hyperedge and joins any number of reactants
to any number of products through a junction point that is stored with the model. Each
reactant and product is joined to the junction by a leg. A reaction can be drawn in
several ways (Fig~\ref{fig:drawing}): with straight
legs meeting at the junction; each leg a cubic B\'ezier curve whose two handles can be
dragged; Bezier curves with a smooth junction, where the curves pass through the junction without the same slope, or for compatibility with Escher reactions~\cite{king2015escher}, with a stem, a straight center line that the reactants join at one end and
the products leave from the other. Reactants and products can be added to or removed from a reaction after it is drawn,
and a side left empty is given a null node. One other style is offered for uni-uni reactions which is a simple straight line.

\begin{figure}[htb]
\centering
\includegraphics[width=\linewidth]{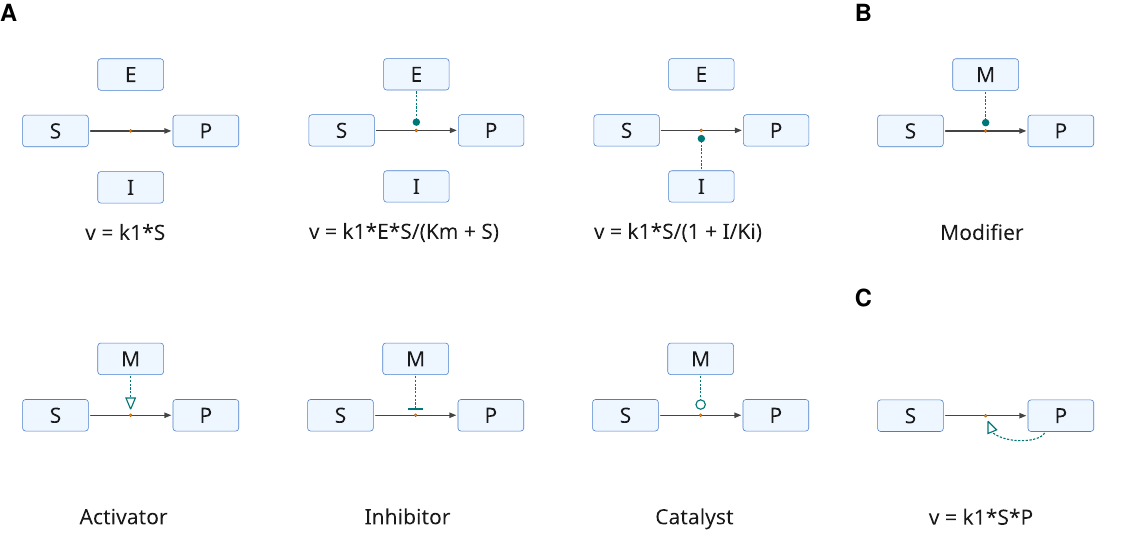}
\caption{{\bf Modifier lines and kinds.} (A) The same reaction after three edits of its
rate law,  with $k_1 S$ there is no modifier line; with a Michaelis--Menten law naming
the enzyme E, a line from E is added; with a law naming an inhibitor I but not E, E's
line is removed and I's added. (B) The four modifier kinds. (C) A product that
activates its own production, marked as a modifier of its reaction and drawn as a loop
from its line back to the junction.}
\label{fig:modifiers}
\end{figure}

\noindent{\bf Modifiers.} Each reaction can have a rate law,
written as an ordinary infix expression such as \texttt{Vm*S/(Km + S)}. As in
JDesigner, modifiers, the species that change a reaction's rate without being used
up, are not drawn by hand. When a rate law is edited, PathwayDesigner reads the species names in the law, and each species  that is not a reactant or product of the reaction becomes a modifier (a rule Antimony also uses), and is automatically connected to the junction by a dashed
modifier line. A modifier that the law no longer names is removed. The modifier lines therefore
reflect the rate laws as they are written (Fig~\ref{fig:modifiers}A). A species that
acts on a rate only through an SBML or Antimony assignment rule is not drawn as a
modifier. A modifier read from an SBML file is kept even if its rate law does not
define it, until that law is next edited.

SBML also allows a reactant or product of a reaction to be listed as a modifier of the same reaction. For example, a product that activates its
own production. Such a self modifier is drawn as a loop from the species' node back to the junction (Fig~\ref{fig:modifiers}C).  For a reaction such as {\tt S -> P} with rate law {\tt k1*S*P}, where {\tt P} also activates the reaction, the user can mark {\tt P} as a modifier with a command on the reaction's right-click menu.

The user can indicate what kind of effect a modifier has, as JDesigner also allowed. It can be drawn as a plain modifier (filled circle), an
activator (an open arrowhead), an inhibitor (a bar) or a catalyst (an open circle)
(Fig~\ref{fig:modifiers}B). A modifier line can be routed around other parts of the
diagram through corner points, as illustrated in the feedback inhibitor in the MAPK
cascade (Fig~\ref{fig:mapk}), and a loop can be turned into corner points when its
default curve is no appropriate, for example it might overlap a species node.

\noindent{\bf Compartments and text.} Cellular compartments are drawn as boxes, and can be
nested. Species are associated with the compartment they reside in. When a compartment is moved the species moves it with.
A compartment's box can be hidden while the compartment, its volume and its species
remain in the model. Free text, for titles and notes, can be placed anywhere on the
canvas.

Full undo and redo is available for all components and alternations of the diagram.

\subsection*{Appearance}

As mentioned previously the appearance of a diagram is independent of the computational model.  Each
species, reaction, compartment and text has its own appearance: fill and border colors,
border width and dash, shape and size, the size, weight and color of its Id, and,
for reactions, the colour and width of the lines. An appearance can be copied from
one item and pasted onto others. It can also be saved under a name as a style and
applied to several items. The items stay linked to the style, so editing the style
changes all of them at once, for example every metabolite in the diagram. Styles are
saved with the diagram in the native json format.

\noindent{\bf Themes} A theme restyles the whole diagram at once. The Classic theme is the default look.
The Escher theme draws metabolites as small orange circles with their names beside
them, smaller circles for cofactors such as ATP and NADH, and thick blue reactions
with their names shown, after the maps drawn by Escher~\cite{king2015escher}. Items
added later take the theme's look. Each modifier line also has a look of its own
(colour, width, dashed, dotted or solid, and the size and placement of its symbol),
which is not part of styles or themes. None of these operations changes the model.

\subsection*{Automatic layout}

A network read from a file without a layout, such as an SBML file or an Antimony
model, has to be laid out. PathwayDesigner uses a
force-directed layout in the style of ForceAtlas2~\cite{jacomy2014forceatlas2},
adapted to hypergraphs: every species node and every reaction junction is a node of
the layout. Every pair of nodes repels, each species is joined by a spring to the
junction of each reaction it takes part in, and a weak gravity holds the network
together. A node's mass grows with the number of legs it has, so that hubs push their
neighbours further away. Species can be locked in place so that they don't move during autolayout. This allows parts of a network to be layout on their own while fixing other parts. The layout algorithm runs a few iterations per frame, so the user can observe the layout being computed. Once it completes, an optional pass sets the B\'ezier handles from the new positions, using the
method of Deckard, Bergmann and Sauro~\cite{deckard2006supporting}. Species and
junctions can also be aligned, distributed and arranged in a circle, and placed on a
grid. This makes it very convenient to make pathway structures such as Krebs's cycle.

\subsection*{Network structure}

Although PathayDesigner has no simulation capability it does have some basic stoichiometric analysis functionality. PathwayDesigner, for example, can display the stoichiometry matrix $\mathbf{N}$. From the stoichiometry matrix $\mathbf{N}$ it can compute the network's conservation laws and a basis for its flux null space using well known published algorithms~\cite{sauro2004conservation}. Both are computing using QR factorization with column
pivoting, as in COPASI~\cite{hoops2006copasi} and libStructural~\cite{vallabhajosyula2006conservation}. From $\mathbf{N}^{\mathsf T}$, the independent species and the link matrix~\cite{reder1988metabolic} can be computed. These provide the conservation laws. From of $\mathbf{N}$ we can compute the independent fluxes and the $\mathbf{K}$ matrix, whose columns are a basis of the null space. A dedicated QR factorization routine was written in Rust for these tasks. Like LAPACK's \texttt{DGEQP3}, which other tools such as COPASI use, our QR algorithm pivots on the largest remaining column norm and ties between columns of equal norm go to
the one listed first in the model.

\begin{figure}[htb]
\centering
\includegraphics[width=\linewidth]{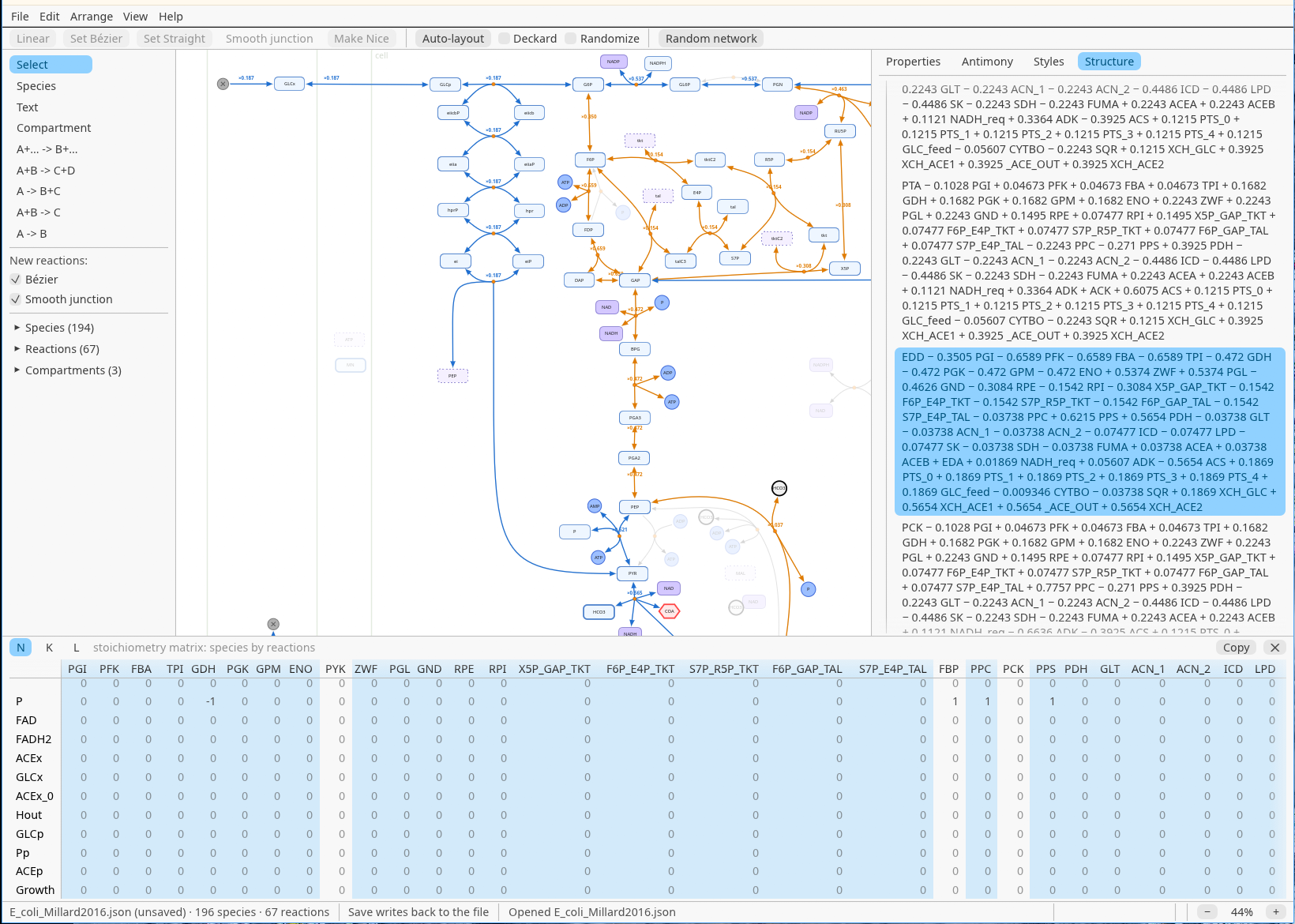}
\caption{{\bf The Structure tab and the matrix panel.} The Millard model~\cite{millard2017metabolic}, with one
vector of the standard null-space basis chosen in the Structure tab (right) and drawn
on the diagram: reactions with a positive coefficient in blue, negative in orange, each
labelled with its coefficient, and the rest of the diagram faded. Below the diagram, can be seen
the stoichiometry matrix, with tabs for the $\mathbf{K}$ and $\mathbf{L}$ matrices.}
\label{fig:structure-ui}
\end{figure}

Both the conservation laws and flux null space can be drawn on to the visible network (Figure~\ref{fig:structure-ui}).

A vector of the null space of $\mathbf{N}$ is a set of reactions whose rates can change together, in
the vector's proportions, while every species steady state concentration remains unchanged~\cite{herbert2019systems}. Because most reaction rates are proportional to the amount of their enzyme, changing the enzymes of the reactions in one such vector in those proportions changes the fluxes along it and leaves the
metabolite concentrations unchanged. The null space is the basis of the
summation theorems of metabolic control analysis and the branch-point
relationships~\cite{fell1985metabolic}.

The basis is however not unique. In the $K$ basis every vector is written in terms of the
dependent reactions the pivoting chose, which on a large network are spread across the network.
This means each vector can overlap several routes. PathwayDesigner therefore also offers a sparse
basis of the same space: starting from the $K$ basis, a vector is replaced by itself
minus the multiple of another that cancels the most entries they share, as long as that
leaves it fewer reactions. Each step makes one vector shorter and the search stops when no single step does, so the result is not always the sparsest basis, which is NP-hard to find in general~\cite{coleman1986null}, but it is deterministic and tends to give short routes. MinSpan~\cite{bordbar2014minimal} finds the sparsest basis for metabolic networks by mixed-integer optimization, but at a  greater cost. The same is done to the conservation laws, so that a law the link matrix gives as the sum of two, such as the ATP and NAD totals, can be split into the two laws.

\begin{figure}[htb]
\centering
\includegraphics[width=\linewidth]{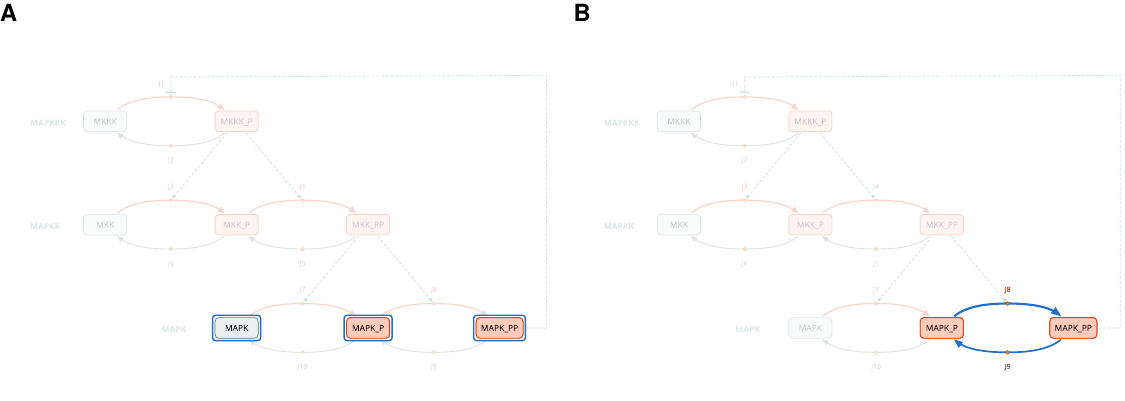}
\caption{{\bf Network structure drawn on the network.} The MAPK cascade with (A) the
conservation law MAPK + MAPK\_P + MAPK\_PP = constant highlighted, the three forms of
MAPK outlined, and (B) the null-space vector J9 + J8, the cycle of phosphorylation and
dephosphorylation of MAPK\_P, highlighted: those two reactions can change together
with every species staying balanced. The rest of the diagram is faded.}
\label{fig:structure}
\end{figure}

\begin{figure}[htb]
\centering
\includegraphics[width=\linewidth]{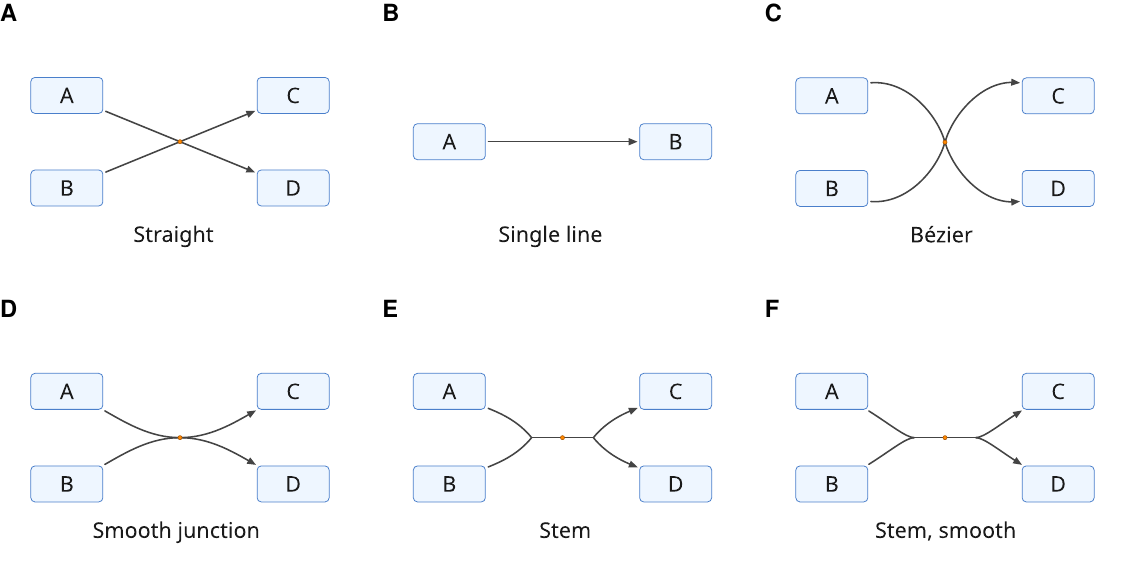}
\caption{The same reaction drawn with straight legs, as a single line, with B\'ezier curves, B\'ezier curves with a smooth junction, B\'ezier curves with a
stem and a stem with smooth B\'ezier curves.}
\label{fig:drawing}
\end{figure}

The results of the stoichiometric analysis are shown in a Structure tab, the rank and the numbers of independent
fluxes and conservation laws, the basis vectors and the laws as equations, and the
stoichiometry matrix in a panel under the diagram. Clicking a vector or a law
highlights it on the network, the rest of the diagram is faded. A vector's reactions
are redrawn in one colour for positive and another for negative coefficients. A law's
species are outlined in every drawing, and coefficients other than 1 are labelled
(Fig~\ref{fig:structure}). The matrix and the lists can be copied or exported as CSV,
and graphics export draws the highlight on the map.

The QR factorization works on the whole matrix, zeros included, so its cost grows steeply with network size. The tab therefore analyses networks of up to about 600 species and 600 reactions, which take about 1.5~s to compute in the browser. Larger networks would need a sparse factorization to be computed efficiently and as a result we currently don't support very large networks for these kinds of calculations (But see~\cite{henderson2016conservation}).

\subsection*{Model and diagram exchange}

\noindent{\bf Native files.} PathwayDesigner can save and load a diagram as a JSON file that stores
both the model and everything about its drawing. The format is versioned for future expansion.

\noindent{\bf Antimony.} Models can be read from and written as
Antimony~\cite{smith2009antimony,heydarabadipour2026antimony3}. The Antimony text is displayed in an edit window where it can be edited. Reading a model into the drawing uses an Antimony reader written for PathwayDesigner in Rust. This library was checked against 1,788 test models and 708 curated BioModels models. PathwayDesigner can read and write Antimony's layout statements~\cite{heydarabadipour2026antimony3}, including positions, lines and looks. An Antimony model that has no layout information is laid out automatically.

\noindent{\bf SBML with Layout and Render.} Models can be read from SBML~\cite{hucka2003systems}
Level~2 and Level~3, and are written as SBML Level~3 Version~2 with the Layout
package~\cite{gauges2006layout,gauges2015layout}, so that a model and its drawing are
held in one standard file. The SBML is read and written by a library written for the
purpose in Rust. Table~\ref{tbl:sbml} lists where each part of a
diagram goes. The drawing's structure goes into the standard Layout package. Modifier
kinds go into the model itself, as Systems Biology Ontology terms~\cite{courtot2011sbo}
on the modifier references (SBO:0000459 for an activator, SBO:0000020 for an inhibitor
and SBO:0000013 for a catalyst), and as the activator and inhibitor roles of Layout.
The diagram's appearance (colors, line widths, shapes, dashes, fonts, arrowheads and
modifier symbols) goes into the Render section~\cite{bergmann2018render}, as one style
for every glyph, keeping to the minimal subset described in Appendix A. Render written by other
tools is read as Appendix A describes. A few drawing details that neither Layout
nor Render expresses, such as a reaction drawn as a single line or a smooth junction,
are kept in short annotations that only PathwayDesigner reads; other tools ignore them. A hidden compartment is written without a box, so that no other
tool draws it, and a compartment that a layout does not draw is read back as hidden.

\begin{table}[htb]
\centering
\caption{{\bf Where each part of a diagram goes in SBML.} Exported as SBML Level 3
Version 2 with the Layout and Render packages.}
\begin{tabular}{p{3.6cm}p{8.6cm}}
\toprule
Where & What \\
\midrule
SBML core & Compartments, species, reactions with their stoichiometry, rate laws and
reversibility, parameters, rules, initial assignments, events, constraints,
functions and units, all as read; modifier kinds as SBO terms \\
Layout package & The place and size of every species and alias, junctions, every
leg's curve, reaction stems, the place of each species' and reaction's name (outside its node
or not), the boxes of shown
compartments, free text; activator and inhibitor roles \\
Render package & Fill, border and line colors and widths, border dashes, species
shapes, fonts (size, bold, italic, color), arrowheads, modifier symbols and their
looks, junction dots, compartment and text styles, the background, saved styles' names \\
PathwayDesigner's annotations, read by PathwayDesigner only & How each reaction is drawn
(single line, smooth junction), hidden compartments,
an alias's automatic border style, text wrapping \\
Not written & The theme (the look of items added later) \\
\bottomrule
\end{tabular}
\label{tbl:sbml}
\end{table}

\noindent{\bf Escher maps.} PathwayDesigner opens the map files of
Escher~\cite{king2015escher} directly. Each metabolite node becomes a species node,
and each further drawing of the same metabolite an alias. Each reaction keeps the
map's own curves and its straight center line as a stem, and the Ids are placed
in the same position as described by Escher. A map opened this way is drawn at Escher's own sizes (lines,
circles and Ids), so that it looks as it does in Escher. The lighter Escher theme,
chosen from the menu, is for smaller diagrams drawn closer together. An Escher map can be saved as
a PathwayDesigner file, or exported as SBML with Layout and Render. Each reaction's gene rule, the genes whose products catalyse it (\texttt{b3916 or b1723}, either gene; \texttt{and} for a complex), is kept and written to SBML with the Flux Balance Constraints package~\cite{olivier2018fbc}.

\noindent{\bf Reactome maps.} The Reactome Knowledgebase~\cite{milacic2024reactome} is an extensive database of annotated biological pathways that can be visualized on their web site. The service offers export of their networks in the SBML Layout format. The SBML unfortunately lacks a layout id which makes the SBML illegal, but the PathwayDesigner code has been adjusted to bypass this issue and assigns a default layout id. This bug has been reported to the Reactome development team. In consequence PathwayDesigner can read, display and edit Reactome networks. An example network of glycolysis is shown in Figure~\ref{fig:reactome}.

\noindent{\bf Images.} A diagram can be exported as SVG, as PNG at any resolution,
and as vector PDF, on a page of a given size or of the drawing's own size. The File menu also offer a print option.

\subsection*{Software Implementation}

PathwayDesigner is written in Rust~\cite{rustlang,Sauro2025Rust}, with about 90,000 lines (that includes comments) in five
libraries (Table~\ref{tbl:crates}). This is in addition to 13,000 lines of test code. The core library holds the model, its geometry
and drawing, the editing commands and their undo, the layout, and every file format.
It contains no user-interface or platform code, so all of it can be tested without a
GUI. The Antimony and SBML libraries are separate, have no dependency on the rest
of PathwayDesigner, and can be used on their own. The user interface is written with
egui~\cite{egui}, an immediate-mode interface library, and drawn through wgpu.

\begin{table}[htb]
\centering
\caption{{\bf The libraries PathwayDesigner is built from.}}
\begin{tabular}{lp{9cm}}
\toprule
Library & Contents \\
\midrule
pd-core & The model, geometry, drawing, editing commands and undo, layout, and the
JSON, Antimony, SBML and Escher bridges. No user interface or platform code. \\
pd-app & The user interface (egui), for the desktop and the browser. \\
pd-export & SVG to PNG conversion. \\
antimony & Reading, checking and writing Antimony, and its conversion to SBML. Pure
Rust, no dependencies. \\
sbml-io & Reading and writing SBML Level 2 and 3 core and Layout, keeping what it does
not interpret. Pure Rust. \\
\bottomrule
\end{tabular}
\label{tbl:crates}
\end{table}

The same code is compiled twice, as a desktop application for Windows, macOS and Linux, and to WebAssembly~\cite{haas2017webassembly} for the browser. This is possible because the entire application is written in pure Rust, that includes the SBML and Antimony support. The browser version also runs entirely in a browser page, with no server. Code that differs between the desktop and browser is confined to a small layer of the user interface: opening and saving files, storing preferences, autosave and printing. Table~\ref{tbl:platforms} lists the differences a user sees.

\begin{table}[htb]
\centering
\caption{{\bf What differs between the desktop and browser versions.}}
\begin{tabular}{>{\raggedright\arraybackslash}p{3.2cm}>{\raggedright\arraybackslash}p{4.2cm}>{\raggedright\arraybackslash}p{4.2cm}}
\toprule
 & Desktop & Browser \\
\midrule
Open & File dialog, the file manager, or the list of recent files & File dialog \\
Save & Back to the opened file & Back to the opened file in Chromium-based browsers
(File System Access API); elsewhere a downloaded copy \\
Preferences, saved styles, autosave & Files in the user's settings folder & The
browser's local storage \\
Recovery after a crash & Offered on the next start & Offered on the next visit \\
Without a network & Everything but the Gallery & Everything but the Gallery, once the
page has been opened \\
\bottomrule
\end{tabular}
\label{tbl:platforms}
\end{table}

The browser version is packaged with Trunk~\cite{trunk}, which compiles the code to WebAssembly and bundles it with the page and its assets as static files that any web server can host. The application compiled into WebAssembly is about 18 MB with its JavaScript
glue, and a service worker that lets it work offline once it has been opened. These
files are served and hosted by a GitHub Page. The application itself runs
in the browser. Any web server can host a copy, as the files need no server-side code.

Every change pushed to the repository has a version tag and automatically builds unsigned installers for Windows, macOS and Linux. Both the desktop and web versions
can open example models from a common Gallery that they read from the project's repository.

Every push is tested on Linux and macOS (and the author's own builds on Windows) with about
1,450 automated tests, the reading and writing of every file in the SBML Test Suite, the validity of every
exported SBML file, and the drawing of PDF files. These tests, and the further checks are described in Appendix~B.

\subsection*{Annotations}

A species can be linked to its entries such as ChEBI~\cite{hastings2016chebi} or UniProt~\cite{uniprot2025}, a reaction to its
EC number or its entry in Rhea~\cite{bansal2022rhea}, and items can have associated Systems Biology
Ontology terms~\cite{courtot2011sbo} and notes, following the MIRIAM
guidelines~\cite{novere2005miriam}. PathwayDesigner preserves these annotations, lets
the user examine and edit them, and can add them to a model that has none. Annotations
that other tools write for their own use, such as COPASI's, are kept and written back
unchanged.

Each item has an \emph{Annotations} part: its SBO term, with its name and a short list of terms that
suit that kind of item. Each reference is a link that opens the entry's page in the
browser through identifiers.org~\cite{juty2012identifiers}. Notes are currently shown as text. A
reference can be added by typing a short identifier such as \texttt{CHEBI:15422}. Hovering the mouse on an item shows its annotation in the
tooltip.

Models can be annotated by hand by looking up each species and reaction. PathwayDesigner can also search for the user, either one item, for a group of selected items or all components. A species is looked up first in BiGG~\cite{king2016bigg}, whose
identifiers are the abbreviations that metabolic models use (\texttt{g6p},
\texttt{accoa}), then in ChEBI by its exact name, through EBI's Ontology Lookup
Service~\cite{jupp2015ols}, and last in PubChem~\cite{kim2023pubchem}. A reaction is looked up in BiGG by its identifier, which gives its EC numbers and its Rhea, KEGG and MetaNetX~\cite{moretti2021metanetx} reactions. Failing that, Rhea is searched by the ChEBI entries of the reaction's species (a species without one is first looked up in BiGG, ChEBI and PubChem). If Rhea finds nothing, BiGG's text search is tried. A gene named in a reaction's gene rule is looked up in UniProt~\cite{uniprot2025} by its locus name (\texttt{b3916}), then by its gene name (\texttt{pfkA}), within the model's organism, which is asked for once and kept with the model. On BiGG's \emph{E. coli} core map~\cite{king2015escher}, 136 of its 137 genes were found, each with a single entry. When PathwayDesigner searches using a short name, the search can return a wrong entry. For example, a search for \texttt{ASP}, meant for aspartate, can also return a peptide. Each match is therefore checked. A species' match is offered as certain only when the entry's own name, not just one of its synonyms, is the name searched for, and a reaction's match only when its participants are the reaction's species.
What was found is shown in a review, with the certain matches ticked and the rest offered with alternatives. A link is given to each
entry and, for a reaction, the equation found to compare with. Until each is resolved, the
items left unticked or with nothing found are marked on the diagram, in orange and
blue. On the \emph{E. coli} model of Millard et al.~\cite{millard2017metabolic}, 17 of its 77 species were offered as certain, all correctly, and then, with those annotated, 31 of its 68 reactions were offered, 7 as certain, again all correctly. Most of the rest are proteins or use abbreviations specific to the model, or are lumped or exchange reactions that no database lists.

\section*{Results}

PathwayDesigner's SBML reader correctly reads and rewrites all 9,350 Level~2 and 3 files of
the SBML Test Suite. Every Gallery model, saved as SBML with Layout and Render, passes libSBML's consistency checks, and COPASI correctly draws these models.
Layouts written by SBMLNetwork, SBMLDiagrams and EscherConverter open correctly in PathwayDesigner. The stoichiometry matrices, conservation laws and null spaces agree with libRoadRunner on all 18 Gallery models and 43 curated BioModels, and the desktop and browser versions give the same output.
Appendix~B gives the details.

\subsection*{Use case 1: an Escher map}

\begin{figure}[htb]
\centering
\includegraphics[width=\linewidth]{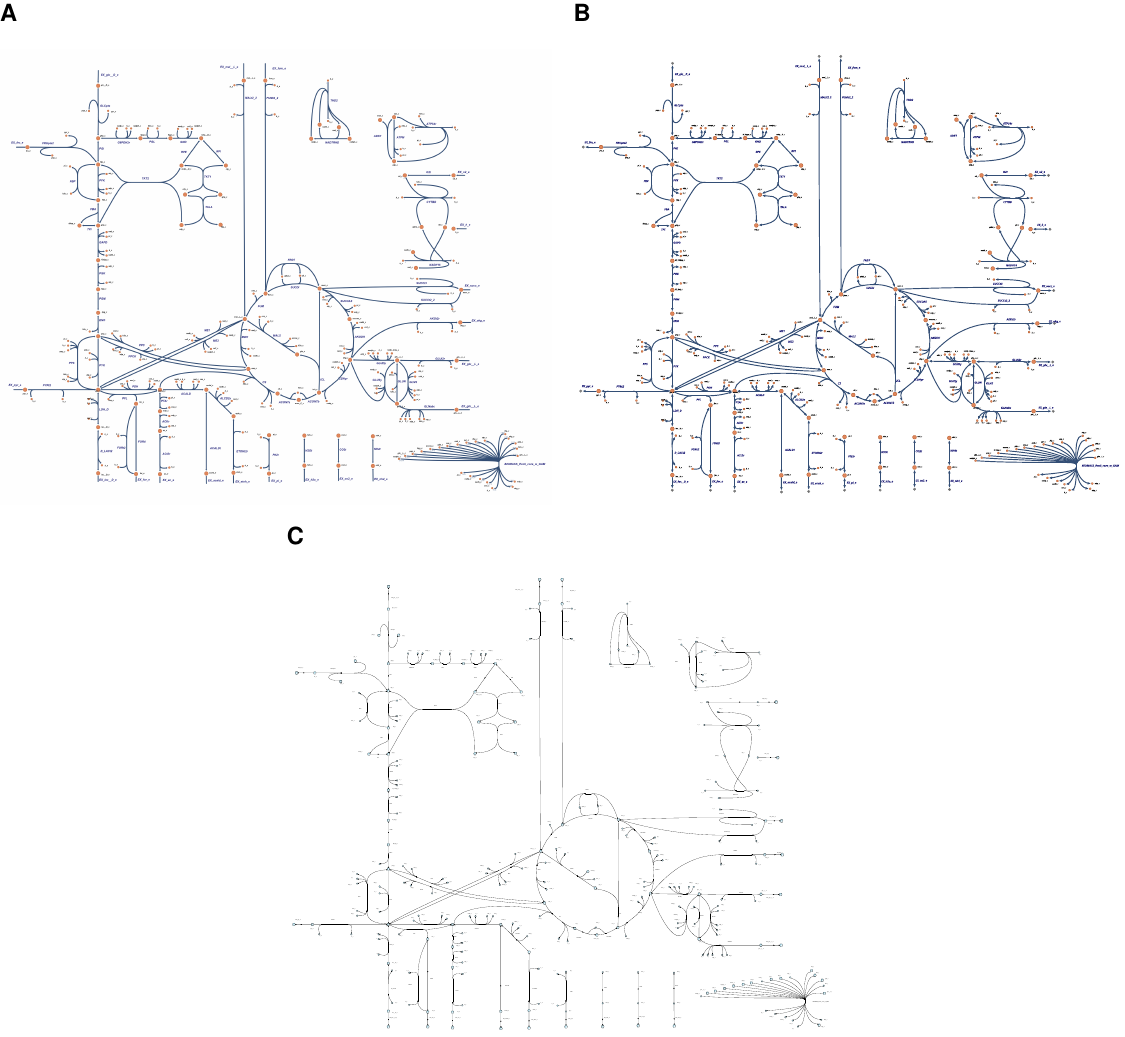}
\caption{{\bf An Escher map, opened and exported by PathwayDesigner.} The Escher map of
\textit{E.~coli} core metabolism drawn by (A) Escher, (B) PathwayDesigner, which reads
the map file directly, and (C) COPASI, from the SBML file with Layout that
PathwayDesigner exports.}
\label{fig:escher}
\end{figure}

Escher~\cite{king2015escher} is a web application for drawing maps of
constraint-based models. Figure~\ref{fig:escher}A, B shows PathwayDesigner rendering of Escher's map of \textit{E.~coli} core. The map's 72 metabolites are drawn 244 times,
so PathwayDesigner reads it as 72 species with 172 aliases, and its 95 reactions with
380 legs maintain the map's original appearance. Because the map is drawn at Escher's own sizes,
the two drawings match closely. Once opened, the map is an
ordinary PathwayDesigner diagram: it can be edited, restyled and given rate laws.
Exported as SBML with Layout, it is a valid SBML file. Once in SBML it can be loaded by a tool such as COPASI which draws it with the
same geometry (Fig~\ref{fig:escher}C). The map's compartments, which Escher does not
draw, are written without boxes, so COPASI does not draw them either.

\begin{figure}[htb]
\centering
\includegraphics[width=\linewidth]{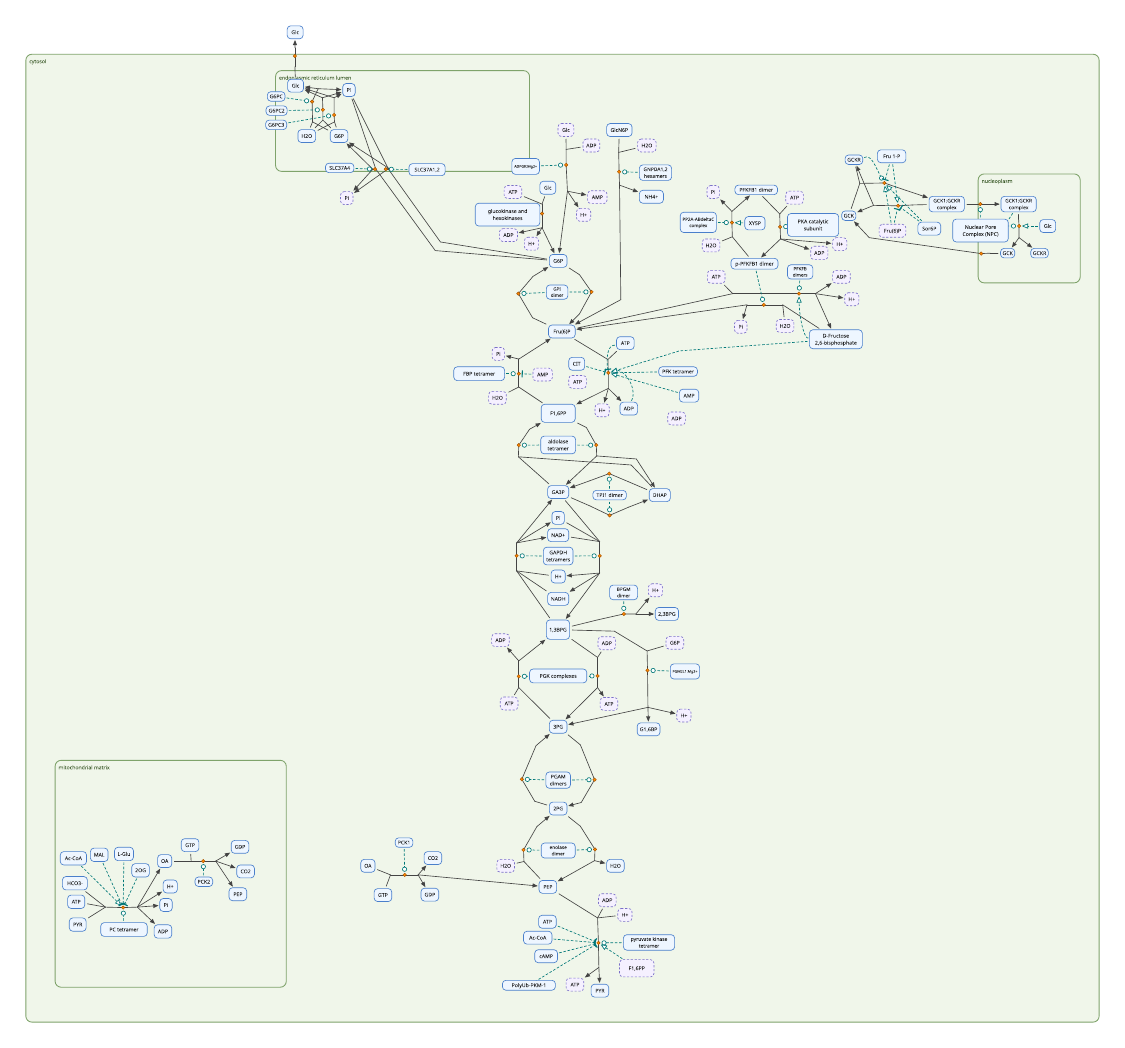}
\caption{{\bf A Reactome pathway opened in PathwayDesigner.} Reactome's glucose
metabolism pathway (R-HSA-70326), from the SBML file Reactome exports, as
PathwayDesigner draws it once opened: Reactome's own layout, its compartments, and the
short labels its layout gives each drawing in place of the full names (\textit{GPI
dimer} for \textit{GPI dimer [cytosol]}). Catalysts and regulators are drawn as
modifier lines of their kinds.}
\label{fig:reactome}
\end{figure}

The same reading was tried on every map in Escher's own map library and on maps drawn
by MetaCarto~\cite{wu2026metacarto}. All 12 maps in the library (five of RECON1, five of iJO1366, and
one each of iMM904 and \textit{E.~coli} core; 1,603 reactions) open without errors or
warnings, with every reaction, every drawing of a metabolite (4,956) and every text
label, and each drawing at the size the map gives it. MetaCarto~\cite{wu2026metacarto}
draws Escher maps of every model in BiGG~\cite{king2016bigg} automatically, and routes
many of a reaction's lines through right-angle bends, which a map drawn in Escher does
not have. Its 96 pathway maps of the human reconstruction
Recon3D~\cite{brunk2018recon3d} (10,598 reactions) open without errors; in about half
the reactions (5,171), lines run through bends, and these are read as 23,173 corner
points on 10,298 lines, one for every bend in the maps, so the routes are drawn as
MetaCarto draws them. MetaCarto leaves water and protons out of its drawings; they are
left out of the reactions concerned. Pathways exported from Reactome as SBML with
Layout also open with their own layout (Fig~\ref{fig:reactome}).

\subsection*{Use case 2: drawing a published kinetic model}

\begin{figure}[htb]
\centering
\includegraphics[width=\linewidth]{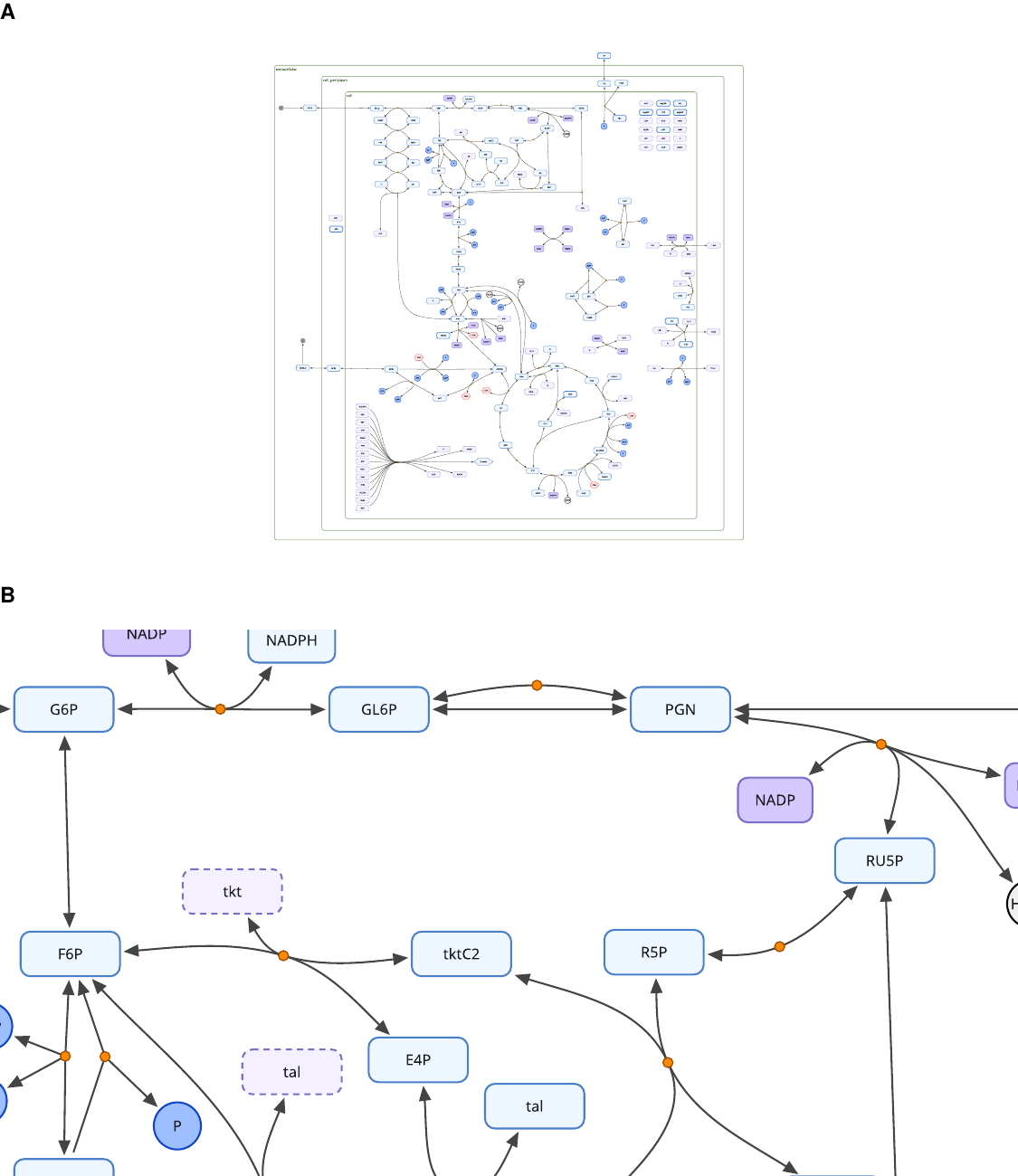}
\caption{{\bf A published kinetic model drawn in PathwayDesigner.} (A) The model of
\textit{E.~coli} central carbon and energy metabolism of Millard et
al.~\cite{millard2017metabolic}, read from its published SBML and laid out by hand
after Fig 1 of that paper. Modifier lines are hidden. (B) Upper glycolysis and the
oxidative pentose phosphate pathway.}
\label{fig:millard}
\end{figure}

Millard, Smallbone and Mendes~\cite{millard2017metabolic} published a kinetic model of
central carbon and energy metabolism in \textit{E.~coli} as SBML (BioModels
MODEL1505110000), with 77 species and 68 reactions but no layout. Their Fig 2 shows the
network. The author opened the published file in PathwayDesigner and laid it out by hand to
reproduce that figure, which took about an hour (Fig~\ref{fig:millard}). The diagram has 156 aliases, 68 reactions and three nested compartments, with one species added, \textit{Growth}, drawn as the product of the growth reaction.
Cofactors such as ATP and NADH,
which take part in many reactions, are drawn as aliases next to the reactions that
use them.

The Structure tab finds the model's 12 conservation laws and 15 independent fluxes in under 2~ms (Fig~\ref{fig:structure-ui}); in the sparse basis a flux vector involves 9 reactions on average, against 35 in the standard basis. The exported SBML, with its layout and render, is valid and can be simulated.

\subsection*{Use case 3: the MAPK cascade with negative feedback}

\begin{figure}[htb]
\centering
\includegraphics[width=\linewidth]{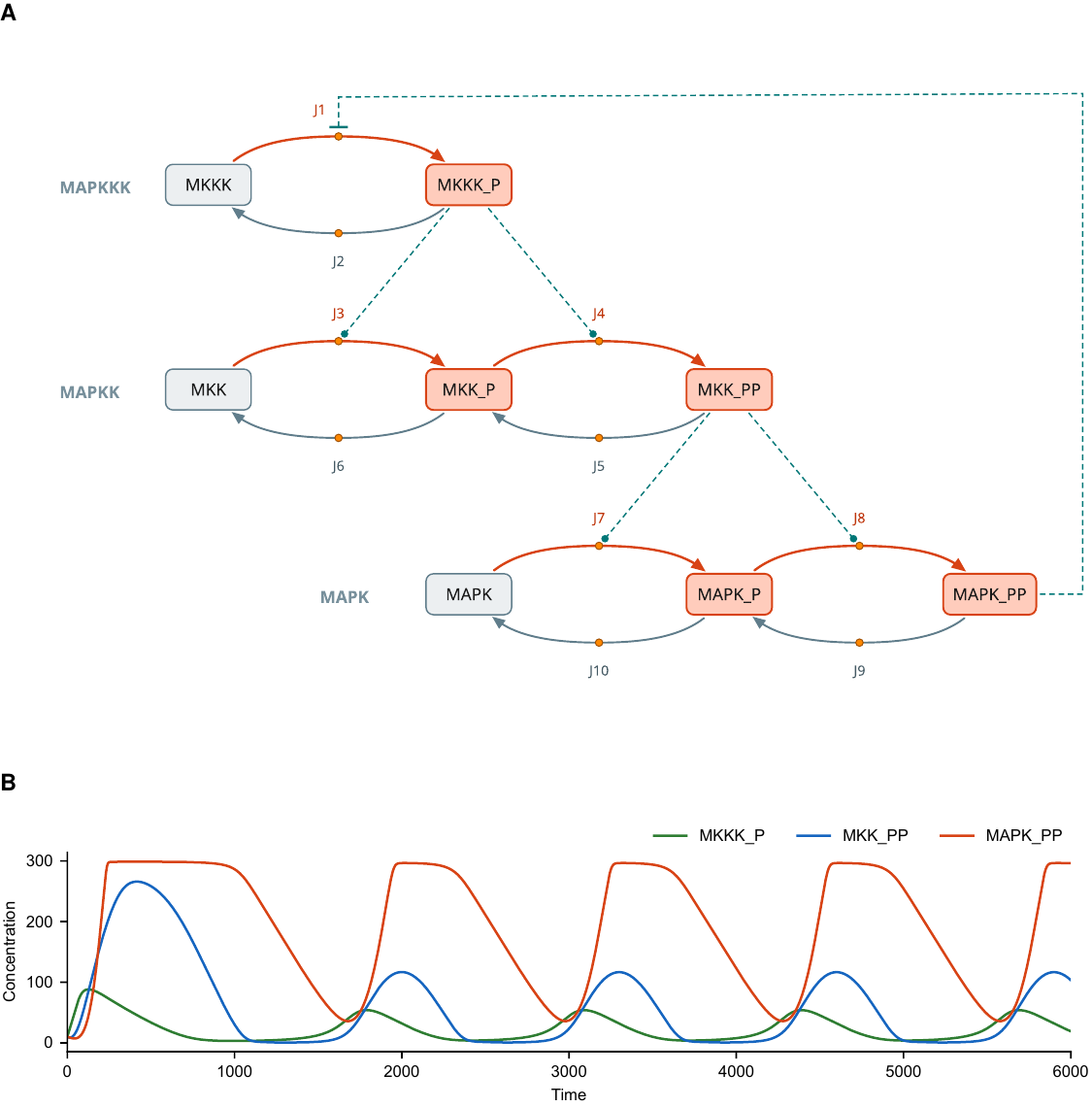}
\caption{{\bf A model and its diagram in one file.} (A) The MAPK cascade with negative
feedback~\cite{kholodenko2000negative} drawn in PathwayDesigner. The five modifier
lines come from the rate laws; MAPK\_PP's line to J1 is an inhibitor, routed round the
diagram through three corner points. (B) The SBML file PathwayDesigner exports for
(A), which also holds its layout, simulated in libRoadRunner.}
\label{fig:mapk}
\end{figure}

The third use case is a small model of the MAPK cascade with negative feedback by Kholodenko~\cite{kholodenko2000negative}
(Fig~\ref{fig:mapk}A). It has 8 species and 10 reactions in three tiers of
phosphorylation and dephosphorylation. Its five modifier lines all come from the rate
laws. The doubly phosphorylated MAPK inhibits the first reaction; its line was set to
the inhibitor kind and routed back round the diagram through three corner points.

The model and its drawing were exported together as one SBML file. Read back, the
file gives the same drawing, including the inhibitor and its route. Simulated in
libRoadRunner, the same file oscillates (Fig~\ref{fig:mapk}B).

\section*{Comparison with other tools}

Table~\ref{tbl:tools} compares PathwayDesigner with other tools that draw biochemical
networks and store the drawing in or beside an SBML model. The entries come from each
tool's documentation, its source code where that was available, and its repository's
activity, all checked in October 2026.

\begin{figure}[htb]
\centering
\includegraphics[width=\linewidth]{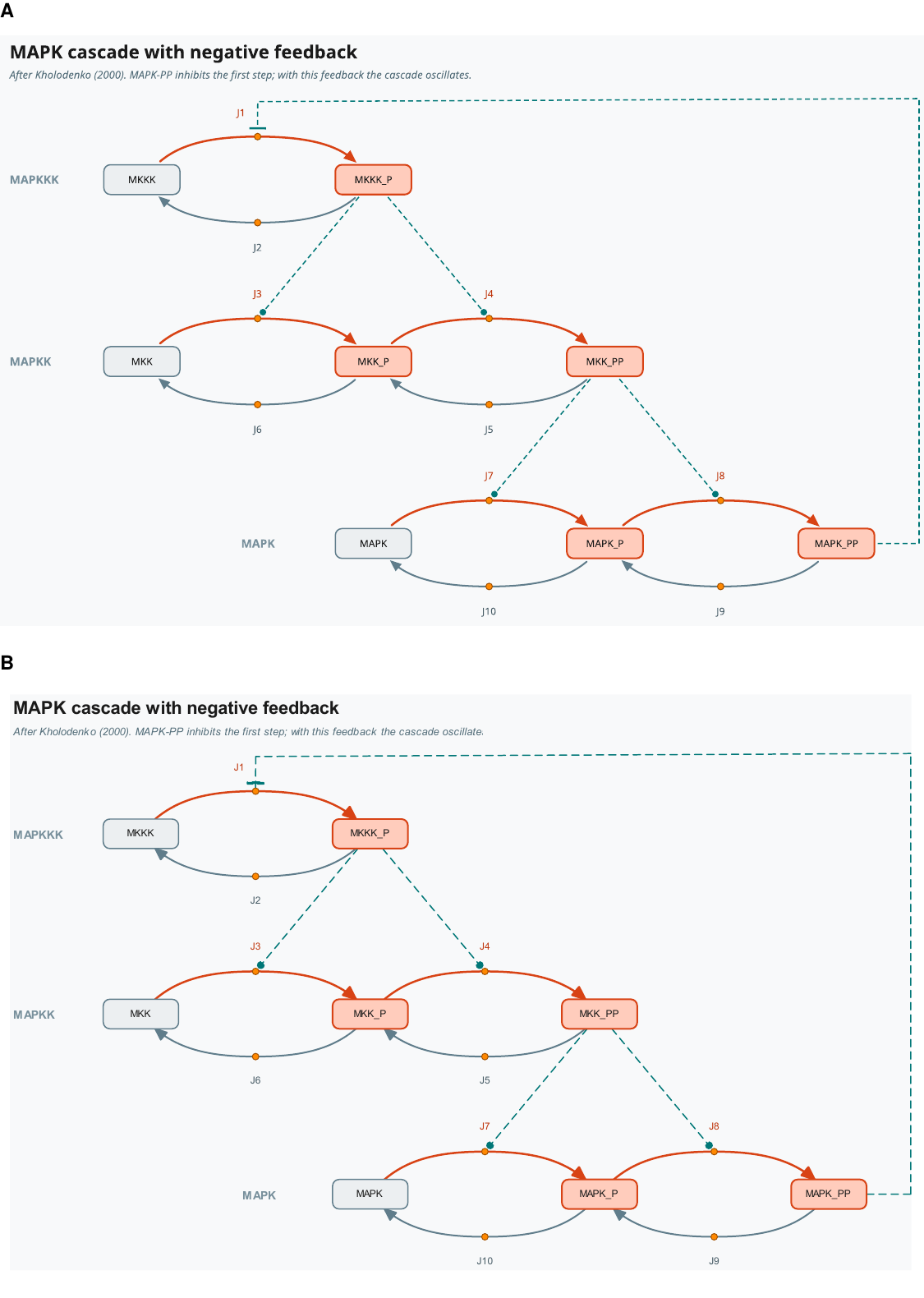}
\caption{{\bf A diagram as drawn SBML Render.} (A) The MAPK cascade drawn by
PathwayDesigner. (B) The SBML file PathwayDesigner exports for it, drawn by COPASI
4.47 from the file's Layout and Render alone. Colours, line widths, shapes, fonts,
junction dots, arrowheads, the inhibitor's bar and the dashed modifier lines come
through.}
\label{fig:copasi}
\end{figure}

\bigskip
\noindent{\bf SBML Layout and Render.} PathwayDesigner is not alone in reading and
writing SBML Layout: COPASI, SBMLDiagrams, SBMLNetwork, Newt and MINERVA do so, and
CellDesigner does through a plugin. COPASI, SBMLDiagrams and SBMLNetwork also read
and write Render, and MINERVA reads part of it. Newt adds an annotation of its own
(\texttt{nwt}) and its SBML converters contain no Render code. Appendix~B shows
what this exchange looks like in practice (Fig~\ref{fig:copasi}).

\noindent{\bf Kinetic model storage.} Of the interactive editors that run in a
browser without installation, Newt, Escher and PathwayDesigner, only PathwayDesigner
keeps a kinetic model with its drawing. Newt's SBML converters carry compartments,
species, reactions, modifiers and the layout, but no rate laws, parameters, initial
values or rules. Escher maps carry none because constraint-based models have none.
The desktop editors JDesigner and CellDesigner keep the model, as do COPASI and the
two libraries.

\noindent{\bf Antimony and Escher.} None of the other tools reads or writes Antimony
as the source of a diagram. However this is understandable given how new the layout format is. SBMLNetwork loads an Antimony model when Tellurium is
installed, and as a library call converts it to SBML. Apart from Escher
itself, none of the tools opens Escher maps. The Java app, EscherConverter, can convert Escher maps to SBML
Layout as a separate step, and SBMLNetwork offers a style template that looks like
Escher's. PathwayDesigner can read an Escher map directly, with its curves and labels.

\noindent{\bf Notation and effort.} Newt, SBGN-ED and MINERVA draw in the Systems
Biology Graphical Notation~\cite{novere2009systems}, and CellDesigner in a notation
close to it. PathwayDesigner, like JDesigner and SBcoyote, draws species as boxes and reactions as lines that meet at a junction. This simplicity makes it possible to draw networks more efficiently due to fewer actions required by the  user to draw a reaction. In one trial, drawing A~$\rightarrow$~B on a blank map in Newt's process-description palette took 10 clicks. In PathwayDesigner it took 6 clicks. For a Bi-Bi reaction newt takes 18 clicks and PathwayDesigner 9 clicks. The difference is entirely due to the added complexity of using SBGN notation. For the Uni-Uni reaction, the SBGN the reaction is three objects in SBGN, a process node and two arcs of different types, and the user has to choose the right type for each arc. Naming the species was not counted in either tool.

\noindent{\bf Network structure.} COPASI computes conservation laws and elementary
flux modes and can show both on its network diagrams~\cite{hoops2006copasi};
CellNetAnalyzer~\cite{klamt2007structural} and its Python version CNApy show
elementary modes and flux distributions on network maps. The number of elementary
modes can grow combinatorially with the size of a network, whereas a basis of the null
space has one vector per independent flux. As far as we found, none of the tools in
Table~\ref{tbl:tools} draws a basis of the flux null space on the network, as
PathwayDesigner does.

\begin{table}[H]
\caption{{\bf Tools that draw biochemical networks.} ``Both, reads and writes'': the
tool reads and writes both SBML Layout and SBML Render. ``Model kept'': the rate laws,
parameters and initial values come back with the drawing. ``Not documented'': not
established from the tool's documentation or code.}
{\footnotesize
\setlength{\tabcolsep}{3pt}
\newcolumntype{L}[1]{>{\raggedright\arraybackslash}p{#1}}
\begin{tabular}{L{2.9cm}L{2.6cm}L{1.2cm}L{2.3cm}L{1.6cm}L{1.5cm}}
\toprule
Tool & What it is, where it runs & Model kept & SBML Layout and Render & Other
formats & Notation \\
\midrule
PathwayDesigner & Interactive editor; browser (static files) and desktop & Yes &
Both, reads and writes (Appendix A) & Antimony (reads, writes); Escher maps (reads) &
Simple \\
JDesigner~\cite{sauro2003next} & Interactive editor; Windows desktop & Yes &
An earlier layout annotation & --- & Simple \\
CellDesigner~\cite{funahashi2008celldesigner} & Interactive editor; desktop (Java) &
Yes & Own annotations; Layout through a plugin & --- & SBGN-like \\
COPASI~\cite{hoops2006copasi} & Simulator that shows and can generate layouts;
desktop & Yes & Both, reads and writes & --- & --- \\
SBcoyote~\cite{xu2023sbcoyote} & Interactive editor; desktop (Python) & Not
documented & Both, reads and writes (SBMLDiagrams) & --- & Simple \\
SBMLDiagrams~\cite{xu2023sbmldiagrams} & Python library, scripted & Yes & Both,
reads and writes & --- & Simple \\
SBMLNetwork~\cite{heydarabadipour2025sbmlnetwork} & Library (C++, Python,
WebAssembly), scripted & Yes & Both, reads and writes & Antimony, through Tellurium &
Templates, SBGN styles \\
Newt~\cite{balci2021newt} & Interactive editor; browser, served by Node.js & No &
Layout only, reads and writes & --- & SBGN \\
SBGN-ED~\cite{czauderna2010sbgned} & Editor and validator for SBGN; desktop (Java,
VANTED) & Not documented & Reads Layout; Render not documented & --- & SBGN \\
MINERVA~\cite{gawron2016minerva} & Viewer for curated maps drawn in other tools
(CellDesigner); web, needs its server & Not documented & Layout; Render partly & --- & SBGN,
CellDesigner \\
Escher~\cite{king2015escher} & Map builder for constraint-based models; browser &
No kinetics & No & Escher maps (native) & Escher's \\
MetaCarto~\cite{wu2026metacarto} & Automatic layout of maps for genome-scale models;
scripted (Python) & No kinetics & No & Models through COBRApy (reads); Escher maps
(writes) & Escher's \\
\bottomrule
\end{tabular}
}
\label{tbl:tools}
\end{table}

\noindent{\bf Maintenance.} JDesigner is no longer maintained and runs only on
Windows. CellDesigner's last release, 4.4.2, dates from 2019. The last changes to
SBcoyote and SBMLDiagrams were made in 2024. COPASI, SBMLNetwork, Newt and Escher
(now developed under the openCOBRA organization) were updated in 2026.

Taken one at a time, each of PathwayDesigner's features is found in at least one
other tool. What the comparison shows is the combination: an interactive editor that
runs in a browser without installation, keeps the kinetic model with its drawing,
reads and writes SBML Layout and Render and Antimony, and can open Escher, KEGG and Reactome maps. It is
closest in purpose to SBMLNetwork, which lays out and styles models from a script;
the two are complementary, since each reads the other's drawings.

\section*{Discussion}

PathwayDesigner is a cross-platform that complements other tools like SBMLNetwork which work via Python scripts.

\subsection*{Limitations}
The annotation lookup only finds metabolites and metabolic reactions, since the services
it searches (BiGG, ChEBI, PubChem and Rhea) hold little else. Proteins are
not looked up in UniProt (but genes are), and a species named with an abbreviation the model's author may have
made up is often not found. For example, a user might use {\tt glco} for glucose.

Annotations that other tools write for their own use, such as COPASI's, are kept and written back unchanged, except those of CellDesigner and JDesigner. These include each tool's own copy of the diagram, which would no longer be true once the diagram is edited in PathwayDesigner, so they are not kept. Standard annotations are always kept, whichever tool wrote them: SBO terms, database references, the model's history and notes.

Some elements of a diagram cannot be exchanged in full. SBML Layout and Render cannot specify whether
a reaction is drawn as a single line or that its junction Beziers are smooth.
PathwayDesigner records these in private SBML annotations, which other tools can ignore. Render is written within the minimal subset of Appendix~A. Other
readers may interpret and draw less. For example, SBMLNetwork does not draw dash arrays or
the background colour. Antimony records a model's modifiers only through its rate laws (its interaction syntax is rarely used), so a modifier that its rate law does not include is lost in an Antimony file.

Imported maps are drawn as well as their files allow. KGML gives no line routes, so
KEGG's curved reactions are approximated by rounded elbows, and KEGG's signalling and
global maps are not read. An SBML file whose layout is incomplete, such as one of the MINERVA maps in Appendix~A, is refused. A model built from submodels (SBML's comp package) is kept and saved whole, but only its top model is shown and edited.

The network structure analysis is limited to about 600 species and reactions. Only one document can be open at a time. It might useful at a future date to add a tab interface to manage multiple projects at the same time. This would be useful to copying parts of networks from one network to another.

A key limitation is we don't support at this time the SBGN standard SBGN-ML.

\subsection*{Future directions}

One significant omission is the ability to render SBGN maps. This is more difficult to achieve since SGBN is more of a visualization standard that a modeling exchange format and PathwayDesigner is specifically designed for simulatable models. WikiPathways is another possible future import, though it also doesn't store models, just the maps.

The network structure analysis could be updated to work on networks greater than about 600 species and reactions. For example, a sparse factorization would make it possible to analyse large genome-scale models (See also~\cite{henderson2016conservation}).

We also plan to add more sophisticated support for searching for and identifying annotations for unannotated models.

\section*{Conclusion}

In this article we describe the development of a new biochemical network editor called PathwayDesigner. This editor is available on all platforms including the Web. The web version can be hosted as a static site on platforms such as GitHub or Cloudflare. PathwayDesigner is the first visual editor that can read and write both SBML Layout and Render standards, though albeit a minimally defined subset of Render that most other tools can read. It can also read and write Antimony, both the layout/render notation and a more concise alternative notation. As well as supporting SBML, the tool can also read the Escher format. It can easily read the SBML Layout format exported by Reactome and can also read KEGG KGML files, although KGML files do not store the reaction geometries so that reactions are rendered as straight lines.

The tool also supports the ability to annotate SBML models such as annotating the species (CheBi, Uniprot, etc) and reactions (such a Rhea or E.C number). For annotating models it has the ability to search responsories for matching names and provide options to the user. For annotating reactions it uses the context of the reaction's reactants and products to search for the name of the reaction. The ability to annotate the top level of an SBML model is also available.

\section*{Availability}
All code and binaries are available from the repository \url{https://github.com/sys-bio/PathwayDesigner-rs} and the web application is available at
\url{https://sys-bio.github.io/PathwayDesigner-rs/}. All code is licensed under the MIT licence. The repo provides build instructions as well as user documentation and a Gallery of sample models.

\section*{Use of AI}
Claude Code (Anthropic, Opus 5.5, medium) was used to assist in writing the software and in preparing this manuscript. Specifically Claude was used to help write the software, debug the code, create the tests and handle submissions of code to the GitHub repo. 1,464 automated tests are run on each build via GitHub continuous integration. In addition,
every file from the SBML Test Suite level 2 and 3 was read and written, and python-libsbml was used to check each result. Round tripping was done on SBML layout files and every test model was passed through libsbml for compliance and consistency. Every reference model was written to Antimony format and read back again to test Antimony compliance. The Antimony implementation was tested against round tripping the 1,788 test models and 708 curated BioModels. Full SBML round tripping was also done on the entire curated branch of the BioModels suite. Annotations were checked on five curated BioModels. PDF output was checked with pypdfium2. All builds run clippy and a licence check. Image comparisons were done for Escher and existing Layout/render files using the SBMLNetwork python package as well as PathwayDesigner-rs itself.

Claude was also used to collate information on standards compliance as well as write sections in Appendix A and B. The original section order in the main text was also devised by Claude and subsequently rearranged and edited by the author. All text was reedited multiple times by the author to ensure accuracy, specificity, and clarity. The author reviewed, verified, and edited all generated content and takes full responsibility for the integrity and accuracy of the final publication. Lessons for AI development as reported for Iridium~\cite{iridium2026} were employed throughout. This section was written by the author.

\section*{Acknowledgments}
This work was supported by NIH Biomedical Imaging and Bioengineering award P41 EB023912 through HMS at the Center for Reproducible Biomedical Modeling (https://reproduciblebiomodels.org/). HMS acknowledges Joe Hellerstein for useful discussions related to AI assistance for application development.

\section*{Appendix A: A proposed minimal subset of SBML Render}

The SBML Render package~\cite{bergmann2018render} can describe almost any drawing, including
groups of primitives, curves, gradients, images, transformations, and styles chosen by
a glyph's id, role or type. Tools that write Render use, however, only a small part of it, and
tools that read it draw only part of what they read. This appendix proposes a minimum
that a writer can keep to and a reader can be expected to draw. This is based on files that
tools actually write, and on what we found when two other tools drew
PathwayDesigner's files. It is offered as a proposal, not a standard.

\subsection*{What tools write}

We surveyed 49 SBML files with Render: 45 written by COPASI 4.45, SBMLNetwork 0.5.11
(including its 24 style templates and the examples from its paper), SBMLDiagrams 1.4.2
and the libSBML samples, and four maps exported by MINERVA 22.0.2~\cite{gawron2016minerva}
from two curated disease maps, two submaps each of the Parkinson's disease map and
the COVID-19 Disease Map. Together they hold 4,027 styles and 722 line endings
(Table~\ref{tbl:survey}). No file used images or transformations, only 9 styles filled
with a gradient, and only MINERVA's used dash arrays (17 styles). 81\% of the styles
were lines and text only, or a single rectangle or ellipse filling the glyph's box;
7\% a single polygon of straight segments, all of them MINERVA's; 4\% a single shape
set inside its box; 1\% the circle and cross of a source or sink; and 7\% several
shapes or curved outlines. These were glyphs of the Systems Biology Graphical
Notation~\cite{novere2009systems} in SBMLNetwork's files and, in MINERVA's, rounded
boxes drawn as polygons with curved sides rather than as rounded rectangles. Line
endings were triangles, open arrows, circles and bars. Most of the styles come from
SBMLNetwork and MINERVA.

\begin{table}[htb]
\centering
\caption{{\bf The styles in 49 SBML files with Render, by what PathwayDesigner can
show of them.} Others: 45 files written by COPASI 4.45, SBMLNetwork 0.5.11,
SBMLDiagrams 1.4.2 and the libSBML samples. MINERVA: four maps exported by MINERVA
22.0.2.}
\setlength{\tabcolsep}{4pt}
\begin{tabular}{p{7.4cm}rrrr}
\toprule
Kind of style & Others & MINERVA & All & Share \\
\midrule
Lines and text only & 1,384 & 922 & 2,306 & 57\% \\
One rectangle or ellipse filling the glyph's box & 763 & 186 & 949 & 24\% \\
One polygon of straight segments filling the box & 0 & 266 & 266 & 7\% \\
One rectangle or ellipse set inside the box & 175 & 0 & 175 & 4\% \\
A circle with a line through it (a source or sink) & 49 & 6 & 55 & 1\% \\
Several shapes, or shapes with curves & 123 & 153 & 276 & 7\% \\
\midrule
All & 2,494 & 1,533 & 4,027 & \\
\bottomrule
\end{tabular}
\label{tbl:survey}
\end{table}

\subsection*{The subset}

Table~\ref{tbl:profile} lists the subset of Render we propose that a tool that writes Render should
use, and what a tool that reads Render should draw. A reader that meets something
outside the subset draws the nearest look it has and informs the user of what it approximated.

\begin{table}[htb]
\centering
\caption{{\bf The proposed minimal Render subset.} Writer: what a tool that writes
Render should use. Reader: what a tool that reads Render should draw; ``approximate''
means draw the nearest look the reader has and tell the user what was approximated;
``honour'' means use the styles as the file chooses them.}
{\small
\newcolumntype{L}[1]{>{\raggedright\arraybackslash}p{#1}}
\begin{tabular}{L{5.2cm}L{2.6cm}L{3.6cm}}
\toprule
Feature & Writer & Reader \\
\midrule
\multicolumn{3}{l}{\it Colours and lines} \\
Colour definitions & Use & Draw \\
Solid fills & Use & Draw \\
Stroke colour and width & Use & Draw \\
Dash arrays & Use & Draw \\
Gradients & Do not use & Draw as one of its colours \\
\multicolumn{3}{l}{\it Text} \\
Font family, size, weight and style & Use & Draw \\
Text anchors & Use & Draw \\
\multicolumn{3}{l}{\it Shapes} \\
One shape filling the glyph's box: rectangle (possibly rounded), ellipse, or polygon
of straight segments & Use & Draw \\
Several shapes in one style; a shape set inside its box; curved polygons & Do not use
& Approximate \\
Images and transformations & Do not use & Approximate \\
Line endings of one shape & Use & Draw \\
\multicolumn{3}{l}{\it Choosing styles} \\
By glyph id, exactly one style per glyph & Use & Honour \\
By role or by type & Do not use & Honour \\
Local render information before the global information it refers to & --- & Honour \\
\bottomrule
\end{tabular}
}
\label{tbl:profile}
\end{table}

\noindent{\bf Writing.} A writer should state every attribute (colour, fill, fill rule,
line width) explicitly on the element it applies to, rather than leaving it to be
inherited from an enclosing group or style, or to a default that a reader may not
implement. The rules for writing the subset are listed below, each with the problem
it avoids:
\begin{enumerate}
\item Give each glyph exactly one style, chosen by its id.
\item Give a fill on each shape, not only on the group that contains it. COPASI draws
a line ending whose fill is given only on its group as an outline.
\item Give a fill rule wherever there is a fill. The specification gives
\texttt{fill-rule} no default, and one of COPASI's renderers fills a shape only when
it has one.
\item Give the colour in every line ending, rather than leaving it to inherit the
line's, so a reader that does not implement the inheritance draws it the same.
\item Write a text's colour as its stroke. Text has a stroke but no fill in Render;
COPASI and SBMLNetwork write it so.
\item This rule is about Layout rather than Render. The rule is to write to the Layout section the curve that is seen on screen. This matters because Render draws line endings, such as arrowheads, at the ends of the Layout curves. A tool may keep a curve in a different form from the one it draws. PathwayDesigner, for example, keeps each curved line as running to the centre of its species, and only cuts it off at the species' border when drawing it. The curve in the SBML Layout should be the visible line, ending at the border, with the control points of that visible piece. Otherwise other tools draw a different curve, with arrowheads pointing the wrong way.
\end{enumerate}

\noindent{\bf Reading.} A reader should accept what writers produce in practice. The
rules for reading the subset are listed below:
\begin{enumerate}
\item Apply inheritance and defaults as the specification gives them, but do not
depend on a writer having used them.
\item Read a text's fill as its colour when it has no stroke.
\item Read tool-defined types before the package's own: SBMLNetwork styles text by
what it labels, with types such as \texttt{SPECIESGLYPH\_TEXTGLYPH}.
\item Read local render information before the global information it refers to.
\item Draw anything outside the subset as the nearest look it has, and tell the user
what was approximated.
\end{enumerate}

\noindent These rules were found by opening PathwayDesigner's exports in COPASI 4.47
and SBMLNetwork 0.5.11: keeping to the features of the subset was not enough for
either to draw the files as PathwayDesigner draws them, and each difference was
resolved by writing the same information in a second, explicit way. The
specification leaves room for such differences: an attribute declared neither on an
element nor in a \texttt{DefaultValues} object is left to the reader, which ``may
chose [sic] how they handle'' it~\cite{bergmann2018render}, and none of the 49 files
in the survey used \texttt{DefaultValues}. Some losses include SBMLNetwork does not draw dash arrays, so dashed borders and modifier
lines appear solid there.

\section*{Appendix B: Validation}

Unless stated otherwise, each check was
run on all the example models in PathwayDesigner's Gallery, which range from a
two-species reaction to the 95-reaction Escher map of \textit{E.~coli} core
metabolism and the 68-reaction kinetic model of Millard et al.~\cite{millard2017metabolic}.
The SBML files were checked with libSBML~\cite{bornstein2008libsbml} (python-libsbml
5.20.5) and simulated with libRoadRunner~\cite{somogyi2015libroadrunner,welsh2023libroadrunner}
(2.9.1).

\subsection*{SBML import and export}

\noindent{\bf Reading and writing SBML.} The SBML library was checked on the semantic
tests of the SBML Test Suite (version 3.5.0), leaving out Level~1: 9,350 files, 6,021
at Level~2 and 3,329 at Level~3. Each file was read and written again. Reading and
writing the result a second time gave the same text for every file. libSBML was then
asked to read the original and the rewritten file and write each out again: the two
outputs were the same for all 9,350 files, and libSBML reported no error in a
rewritten file that it did not report in the original. The Test Suite contains no
layouts, so the two SBML Layout example files are checked separately. Both checks run
on every change to the repository.

Every Gallery model, the Escher map included, was exported as SBML with Layout and Render and checked with libSBML's consistency checks. No errors were detected. libSBML has no validation rules for Render, so the Render part was checked by reading and drawing the files in other tools (see Appendix A).

The published SBML of the Millard model defines 61 functions, which its 68 rate laws call. PathwayDesigner keeps them as written; with its preference to write functions into the formulas that use them, the exported file has none. Simulated from the same starting point, the published file and both exports give identical values for all 64 floating species.

\noindent{\bf Round trip through SBML.} Each Gallery model was exported as SBML with Layout and Render and read back, and the result was compared with the original. Everything drawn came back. Two things are not written: the theme, which decides how items added later will look (3 models), and the boxes of hidden compartments (two in the Escher map, three in the Reactome model). The remaining differences do not change the drawing.

\noindent{\bf Models from other tools.} The 1,062 curated BioModels models and the
1,823 semantic cases of the SBML Test Suite were each read by PathwayDesigner and
written back as SBML. Compared element by element with libSBML's own conversion of the
original to the same SBML level and version, formulas by value, every file came back
unchanged, events, rules, constraints, functions, units and local parameters
included. So did every file written as Antimony with its diagram and read back.
Simulated with libRoadRunner before and after, 1,033 of the BioModels models behaved
the same (918 bit for bit); 28 did not simulate to begin with, and the one remaining
difference, about $10^{-12}$ in absolute terms, comes from the solver. Of the Test
Suite cases that libRoadRunner itself passes, all but four passed after the round
trip; those four fail in libRoadRunner when libSBML converts the original too.

\noindent{\bf Round tripping through Antimony.} Each Gallery model was exported to Antimony, with its diagram, and read back. Species, initial values, boundary and constant species, compartments, stoichiometry, reversibility, rate laws, parameters, rules, and the modifiers and their kinds came back unchanged in every model, the drawing included. What Antimony cannot say, such as a reaction without a rate law or a name that Antimony reserves, is kept in a second comment after the diagram. Written as plain Antimony, without the diagram, these are lost, and PathwayDesigner says which. Exported as SBML without a layout, the same models came back unchanged, apart from the name of seven models that had none, which take the model's id.

\subsection*{Reading layouts written by other tools}

COPASI~\cite{hoops2006copasi} reads the SBML Layout and Render that PathwayDesigner writes and
draws it with the same geometry (Fig~\ref{fig:escher}C). In the
other direction, three tools that write SBML Layout and Render were used to lay out models, and
PathwayDesigner read the results (Table~\ref{tbl:otherlayouts}): SBMLNetwork
0.5.11~\cite{heydarabadipour2025sbmlnetwork} and SBMLDiagrams 1.4.2~\cite{xu2023sbmldiagrams}, each on the MAPK
cascade and the Millard model, and EscherConverter 1.2.1~\cite{EscherConverter2021} on the Escher map of
\textit{E.~coli} core metabolism. Each species glyph, as libSBML read it, was matched
to the node PathwayDesigner made for it. The Render information these tools
write is read as Appendix A describes. What is not carried over are the curves of
modifier lines, which PathwayDesigner draws straight or through corner points.

\noindent{\bf Render read and written by other tools.} PathwayDesigner read all 45
of the survey's files from COPASI, SBMLNetwork, SBMLDiagrams and libSBML (Appendix A)
without error. Only the two SBMLNetwork
examples drawn in the Systems Biology Graphical Notation, with glyphs of several
shapes (18 in all), and three of libSBML's samples, with gradients, needed
approximating, each with one warning. The files of COPASI, SBMLDiagrams and
SBMLNetwork's other examples came through exactly. Of the four MINERVA maps added
to the survey, three were read, each with one warning for its rounded boxes drawn as
curved polygons; the fourth was refused, because MINERVA writes a second layout, for a
hidden layer, without the dimensions SBML Layout requires (libSBML's consistency check
reports the same error). MINERVA's own drawing of a compartment also differs from its
export: it draws a thick border round a pale inside, but writes the border's colour as
an opaque fill, which PathwayDesigner draws as indicated in the file. In the other direction, COPASI
4.47 and SBMLNetwork 0.5.11 drew PathwayDesigner's exports in its colours, shapes,
fonts, junction dots, arrowheads and modifier symbols, once the writing rules of
Appendix A were followed (Fig~\ref{fig:copasi}).

\begin{table}[htb]
\centering
\caption{{\bf Layouts written by other tools, read by PathwayDesigner.} Glyphs that
name no species become null nodes. Worst offset: the largest distance between a
glyph's centre and the centre of the node made for it.}
\begin{tabular}{llrrr}
\toprule
Layout written by & Model & \shortstack[r]{Species\\glyphs} & \shortstack[r]{Became\\aliases} & \shortstack[r]{Worst\\offset} \\
\midrule
SBMLNetwork 0.5.11 & MAPK cascade & 12 & 4 & 0 \\
SBMLNetwork 0.5.11 & Millard et al. & 149 & 69 & 0 \\
SBMLDiagrams 1.4.2 & MAPK cascade & 8 & 0 & 0 \\
SBMLDiagrams 1.4.2 & Millard et al. & 78 & 0 & 0 \\
EscherConverter 1.2.1 & \textit{E.~coli} core map & 244 & 172 & 0 \\
\bottomrule
\end{tabular}
\label{tbl:otherlayouts}
\end{table}

PathwayDesigner also opens Escher map files directly. For the \textit{E.~coli} core
map, its reading and EscherConverter's agree: the same 72 species, 172 aliases, 95
reactions and 380 legs, with every node and junction in the same place once
EscherConverter's shift of the whole map to its origin is allowed for.

\subsection*{Network structure checked against libRoadRunner}

The stoichiometry matrix, rank, conservation laws and null-space basis were compared,
model by model, with those of libRoadRunner~\cite{welsh2023libroadrunner}. For SBML
models libRoadRunner read the original file, so PathwayDesigner's SBML import was
checked too. The matrices, entry by entry, and the ranks agreed in all 18 Gallery
models and in 43 of 45 curated BioModels; the other two contain delays, which
libRoadRunner does not load. In each of these models every PathwayDesigner basis vector
$\mathbf{v}$ satisfied $\mathbf{N} \mathbf{v} = \mathbf{0}$, the null-space basis spanned the same space as libRoadRunner's
$\mathbf{K}$ matrix, and the conservation laws spanned the same space as its conservation
matrix; the sparse bases spanned the same spaces too. Seven of the BioModels declare
species constant that act only as modifiers; PathwayDesigner leaves these out of $N$,
as they cannot change, where libRoadRunner keeps them as rows of zeros, and they were
dropped from its side for the comparison. In five, libRoadRunner declined its moiety
analysis (species set by rules, or variable stoichiometry), and the rank and laws were
checked against a singular value decomposition instead. Which species are independent
agreed in about two thirds of the models; in the rest PathwayDesigner chose another,
equally valid set, because ties between columns of equal norm are broken differently.

The sparse basis shortened the vectors: the mean number of reactions per vector fell
from 6.2 to 3.9 over the BioModels, from 40 to 15.9 on the \textit{E.~coli} core map, and
from 34.7 to 6.5 (the longest from 69 to 38) on the RECON1 carbohydrate map of
Escher's library, 366 species by 340 reactions, which was analysed in 91~ms on the
desktop and 207~ms in the browser, and its sparse basis found in 95~ms on the desktop.
Fig~\ref{fig:structure} shows a conservation law and a null-space vector of the MAPK
cascade highlighted.

Fig~\ref{fig:structure-ui} shows the same tools on a larger network, the Millard model
of Use case~2: 65 floating species and 68 reactions, rank 53, with 12 conservation
laws and a null space of dimension 15, found in under 2~ms on the desktop. The
Structure tab lists the basis vectors and the laws; clicking one draws it on the
diagram and fades the rest, and the panel below the diagram shows the stoichiometry
matrix, with tabs for $\mathbf{K}$ and $\mathbf{L}$. In the standard basis a vector of this model involves
35 reactions on average (at most 47); in the sparse basis, 9.3 (at most 30).

\subsection*{Desktop and browser builds compared}

The core library was compiled for the desktop and for WebAssembly, and both builds
were given the same task for each Gallery model: save it, export it as Antimony and
SBML and read each back, draw it as SVG and PDF, and lay it out automatically from
the same random start. The saved files, the Antimony, the SVG and the PDF were the
same, byte for byte, in every model. The SBML and the automatic layouts differed only
in the last digit of a few numbers, because the two builds compute the length of a
vector with different system maths libraries; with that one function written out in
full, every output was the same.

\bibliography{refs,refs_pd}

@article{hucka2003systems,
  title={The systems biology markup language (SBML): a medium for representation and exchange of biochemical network models},
  author={Hucka, Michael and Finney, Andrew and Sauro, Herbert M and Bolouri, Hamid and Doyle, John C and Kitano, Hiroaki and Arkin, Adam P and Bornstein, Benjamin J and Bray, Dennis and Cornish-Bowden, Athel and others},
  journal={Bioinformatics},
  volume={19},
  number={4},
  pages={524--531},
  year={2003},
  publisher={Oxford University Press},
  doi={10.1093/bioinformatics/btg015}
}

@article{sauro2003next,
  title={Next generation simulation tools: the Systems Biology Workbench and BioSPICE integration},
  author={Sauro, Herbert M and Hucka, Michael and Finney, Andrew and Wellock, Cameron and Bolouri, Hamid and Doyle, John and Kitano, Hiroaki},
  journal={Omics: A journal of integrative biology},
  volume={7},
  number={4},
  pages={355--372},
  year={2003},
  publisher={SAGE Publications Sage CA: Los Angeles, CA}
}

@article{bergmann2018sbml,
  title={SBML level 3 package: render, version 1, release 1},
  author={Bergmann, Frank T and Keating, Sarah M and Gauges, Ralph and Sahle, Sven and Wengler, Katja},
  journal={Journal of Integrative Bioinformatics},
  volume={15},
  number={1},
  pages={20170078},
  year={2018}
}

@article{gauges2015systems,
  title={The systems biology markup language (SBML) level 3 package: layout, version 1 core},
  author={Gauges, Ralph and Rost, Ursula and Sahle, Sven and Wengler, Katja and Bergmann, Frank Thomas},
  journal={Journal of Integrative Bioinformatics},
  volume={12},
  number={2},
  year={2015}
}

@article{hoops2006copasi,
  title={COPASI—a complex pathway simulator},
  author={Hoops, Stefan and Sahle, Sven and Gauges, Ralph and Lee, Christine and Pahle, J{\"u}rgen and Simus, Natalia and Singhal, Mudita and Xu, Liang and Mendes, Pedro and Kummer, Ursula},
  journal={Bioinformatics},
  volume={22},
  number={24},
  pages={3067--3074},
  year={2006},
  publisher={Oxford University Press}
}

@article{welsh2023libroadrunner,
  title={libRoadRunner 2.0: a high performance SBML simulation and analysis library},
  author={Welsh, Ciaran and Xu, Jin and Smith, Lucian and K{\"o}nig, Matthias and Choi, Kiri and Sauro, Herbert M},
  journal={Bioinformatics},
  volume={39},
  number={1},
  pages={btac770},
  year={2023},
  publisher={Oxford University Press}
}

@article{funahashi2008celldesigner,
  title={CellDesigner 3.5: a versatile modeling tool for biochemical networks},
  author={Funahashi, Akira and Matsuoka, Yukiko and Jouraku, Akiya and Morohashi, Mineo and Kikuchi, Norihiro and Kitano, Hiroaki},
  journal={Proceedings of the IEEE},
  volume={96},
  number={8},
  pages={1254--1265},
  year={2008},
  publisher={Ieee}
}

@misc{EscherConverter2021,
  author       = {Andreas Dräger},
  title        = {Escher Converter},
  year         = {2021},
  url          = {https://github.com/draeger-lab/EscherConverter}
}

@article{smith2009antimony,
  title={Antimony: a modular model definition language},
  author={Smith, Lucian P and Bergmann, Frank T and Chandran, Deepak and Sauro, Herbert M},
  journal={Bioinformatics},
  volume={25},
  number={18},
  pages={2452--2454},
  year={2009},
  publisher={Oxford University Press}
}

@article{novere2009systems,
  title={The systems biology graphical notation},
  author={Nov{\`e}re, Nicolas Le and Hucka, Michael and Mi, Huaiyu and Moodie, Stuart and Schreiber, Falk and Sorokin, Anatoly and Demir, Emek and Wegner, Katja and Aladjem, Mirit I and Wimalaratne, Sarala M and others},
  journal={Nature biotechnology},
  volume={27},
  number={8},
  pages={735--741},
  year={2009},
  publisher={Nature Publishing Group US New York},
  doi={10.1038/nbt.1558}
}

@article{xu2023sbcoyote,
  author    = {Xu, Jin and Geng, Gary and Nguyen, Nhan D. and Perena-Cortes, Carmen and Samuels, Claire and Sauro, Herbert M.},
  title     = {SBcoyote: An extensible Python-based reaction editor and viewer},
  journal   = {Biosystems},
  volume    = {232},
  pages     = {105001},
  year      = {2023},
  month     = {October},
  doi       = {10.1016/j.biosystems.2023.105001},
  publisher = {Elsevier}
}

@article{somogyi2015libroadrunner,
  title={libRoadRunner: a high performance SBML simulation and analysis library},
  author={Somogyi, Endre T and Bouteiller, Jean-Marie and Glazier, James A and K{\"o}nig, Matthias and Medley, J Kyle and Swat, Maciej H and Sauro, Herbert M},
  journal={Bioinformatics},
  volume={31},
  number={20},
  pages={3315--3321},
  year={2015},
  publisher={Oxford University Press}
}

@article{gauges2006layout,
  author  = {Gauges, Ralph and Rost, Ursula and Sahle, Sven and Wegner, Katja},
  title   = {A model diagram layout extension for {SBML}},
  journal = {Bioinformatics},
  volume  = {22},
  number  = {15},
  pages   = {1879--1885},
  year    = {2006},
  doi     = {10.1093/bioinformatics/btl195}
}

@article{gauges2015layout,
  author  = {Gauges, Ralph and Rost, Ursula and Sahle, Sven and Wengler, Katja and Bergmann, Frank T.},
  title   = {The {Systems Biology Markup Language (SBML)} Level 3 Package: Layout, Version 1 Core},
  journal = {Journal of Integrative Bioinformatics},
  volume  = {12},
  number  = {2},
  pages   = {550--602},
  year    = {2015},
  doi     = {10.1515/jib-2015-267}
}

@article{deckard2006supporting,
  author  = {Deckard, Anastasia and Bergmann, Frank T. and Sauro, Herbert M.},
  title   = {Supporting the {SBML} layout extension},
  journal = {Bioinformatics},
  volume  = {22},
  number  = {23},
  pages   = {2966--2967},
  year    = {2006},
  doi     = {10.1093/bioinformatics/btl520}
}

@article{bergmann2018render,
  author  = {Bergmann, Frank T. and Keating, Sarah M. and Gauges, Ralph and Sahle, Sven and Wengler, Katja},
  title   = {{SBML} Level 3 package: Render, Version 1, Release 1},
  journal = {Journal of Integrative Bioinformatics},
  volume  = {15},
  number  = {1},
  year    = {2018},
  doi     = {10.1515/jib-2017-0078}
}

@article{king2015escher,
  author  = {King, Zachary A. and Dr{\"a}ger, Andreas and Ebrahim, Ali and Sonnenschein, Nikolaus and Lewis, Nathan E. and Palsson, Bernhard O.},
  title   = {Escher: A Web Application for Building, Sharing, and Embedding Data-Rich Visualizations of Biological Pathways},
  journal = {PLOS Computational Biology},
  volume  = {11},
  number  = {8},
  pages   = {e1004321},
  year    = {2015},
  doi     = {10.1371/journal.pcbi.1004321}
}

@article{heydarabadipour2025sbmlnetwork,
  author  = {Heydarabadipour, Adel and Smith, Lucian and Hellerstein, Joseph L. and Sauro, Herbert M.},
  title   = {{SBMLNetwork}: A framework for standards-based visualization of biochemical models},
  journal = {PLOS Computational Biology},
  volume  = {21},
  number  = {9},
  pages   = {e1013128},
  year    = {2025},
  doi     = {10.1371/journal.pcbi.1013128}
}

@article{heydarabadipour2026antimony3,
  author  = {Heydarabadipour, Adel and Smith, Lucian P. and Hellerstein, Joseph L. and Sauro, Herbert M.},
  title   = {Antimony 3: Extending human-readable model definitions for {SBML} Level 3 Core and Packages},
  journal = {PLOS Computational Biology},
  volume  = {22},
  number  = {10},
  pages   = {e1014207},
  year    = {2026},
  doi     = {10.1371/journal.pcbi.1014207}
}

@article{jacomy2014forceatlas2,
  author  = {Jacomy, Mathieu and Venturini, Tommaso and Heymann, Sebastien and Bastian, Mathieu},
  title   = {{ForceAtlas2}, a Continuous Graph Layout Algorithm for Handy Network Visualization Designed for the {Gephi} Software},
  journal = {PLoS ONE},
  volume  = {9},
  number  = {6},
  pages   = {e98679},
  year    = {2014},
  doi     = {10.1371/journal.pone.0098679}
}

@article{courtot2011sbo,
  author  = {Courtot, M{\'e}lanie and Juty, Nick and Kn{\"u}pfer, Christian and Waltemath, Dagmar and others},
  title   = {Controlled vocabularies and semantics in systems biology},
  journal = {Molecular Systems Biology},
  volume  = {7},
  number  = {1},
  pages   = {543},
  year    = {2011},
  doi     = {10.1038/msb.2011.77}
}

@article{coleman1986null,
  title={The null space problem I. Complexity},
  author={Coleman, Thomas F and Pothen, Alex},
  journal={SIAM Journal on Algebraic Discrete Methods},
  volume={7},
  number={4},
  pages={527--537},
  year={1986},
  publisher={SIAM},
  doi={10.1137/0607059}
}

@article{milacic2024reactome,
  title={The reactome pathway knowledgebase 2024},
  author={Milacic, Marija and Beavers, Deidre and Conley, Patrick and Gong, Chuqiao and Gillespie, Marc and Griss, Johannes and Haw, Robin and Jassal, Bijay and Matthews, Lisa and May, Bruce and others},
  journal={Nucleic acids research},
  volume={52},
  number={D1},
  pages={D672--D678},
  year={2024},
  publisher={Oxford University Press},
  doi={10.1093/nar/gkad1025}
}

@article{bordbar2014minimal,
  title={Minimal metabolic pathway structure is consistent with associated biomolecular interactions},
  author={Bordbar, Aarash and Nagarajan, Harish and Lewis, Nathan E and Latif, Haythem and Ebrahim, Ali and Federowicz, Stephen and Schellenberger, Jan and Palsson, Bernhard O},
  journal={Molecular systems biology},
  volume={10},
  number={7},
  pages={737},
  year={2014},
  publisher={Springer},
  doi={10.15252/msb.20145243}
}

@article{haas2017webassembly,
  author  = {Haas, Andreas and Rossberg, Andreas and Schuff, Derek L. and Titzer, Ben L. and Holman, Michael and Gohman, Dan and Wagner, Luke and Zakai, Alon and Bastien, J. F.},
  title   = {Bringing the web up to speed with {WebAssembly}},
  journal = {ACM SIGPLAN Notices},
  volume  = {52},
  number  = {6},
  pages   = {185--200},
  year    = {2017},
  doi     = {10.1145/3140587.3062363}
}

@article{millard2017metabolic,
  author  = {Millard, Pierre and Smallbone, Kieran and Mendes, Pedro},
  title   = {Metabolic regulation is sufficient for global and robust coordination of glucose uptake, catabolism, energy production and growth in {Escherichia coli}},
  journal = {PLOS Computational Biology},
  volume  = {13},
  number  = {2},
  pages   = {e1005396},
  year    = {2017},
  doi     = {10.1371/journal.pcbi.1005396}
}

@misc{egui,
  author       = {Ernerfeldt, Emil and contributors},
  title        = {egui: an immediate mode {GUI} library for {Rust}},
  howpublished = {\url{https://github.com/emilk/egui}},
  note         = {Version 0.37},
  year         = {2026}
}

@misc{rustlang,
  author       = {{The Rust Project Developers}},
  title        = {The {Rust} Programming Language},
  howpublished = {\url{https://www.rust-lang.org}},
  year         = {2026}
}

@book{Sauro2025Rust,
author    = {Herbert M. Sauro},
title     = {Rust for Beginners},
publisher = {Ambrosius Publishing},
year      = {2025},
isbn      = {978-1732548671},
month     = {June},
note      = {Paperback, 112 pages}
}

@misc{trunk,
  author       = {{Trunk contributors}},
  title        = {Trunk: build, bundle and ship your {Rust} {WebAssembly} application to the web},
  howpublished = {\url{https://github.com/trunk-rs/trunk}},
  year         = {2026}
}

@misc{iridium2026,
  author        = {Sauro, Herbert M. and Nguyen, Dan and Nguyenle, Long and Andrews, Steven S. and Bergmann, Frank T. and Hellerstein, Joseph L. and Smith, Lucian P. and Wiley, H. Steven},
  title         = {An interactive simulator for integrating biochemical models with experimental data},
  year          = {2026},
  eprint        = {2609.34186},
  archivePrefix = {arXiv},
  primaryClass  = {q-bio.MN},
  howpublished  = {arXiv:2609.34186}
}

@article{bornstein2008libsbml,
  author  = {Bornstein, Benjamin J. and Keating, Sarah M. and Jouraku, Akiya and Hucka, Michael},
  title   = {{LibSBML}: an {API} Library for {SBML}},
  journal = {Bioinformatics},
  volume  = {24},
  number  = {6},
  pages   = {880--881},
  year    = {2008},
  doi     = {10.1093/bioinformatics/btn051}
}

@article{kholodenko2000negative,
  author  = {Kholodenko, Boris N.},
  title   = {Negative feedback and ultrasensitivity can bring about oscillations in the mitogen-activated protein kinase cascades},
  journal = {European Journal of Biochemistry},
  volume  = {267},
  number  = {6},
  pages   = {1583--1588},
  year    = {2000},
  doi     = {10.1046/j.1432-1327.2000.01197.x}
}

@article{sauro2004conservation,
  title={Conservation analysis in biochemical networks: computational issues for software writers},
  author={Sauro, Herbert M and Ingalls, Brian},
  journal={Biophysical chemistry},
  volume={109},
  number={1},
  pages={1--15},
  year={2004},
  publisher={Elsevier},
  doi={10.1016/j.bpc.2003.08.009}
}

@misc{PathwayDesignerPrototype,
  author       = {Sauro, Herbert M.},
  title        = {Desktop PathwayDesigner: A Prototype},
  howpublished = {https://github.com/hsauro/PathwayDesigner},
  year         = {2026}
}

@article{balci2021newt,
  title={Newt: a comprehensive web-based tool for viewing, constructing and analyzing biological maps},
  author={Balci, Hasan and Siper, Metin Can and Saleh, Nasim and Safarli, Ilkin and Roy, Ludovic and Kilicarslan, Merve and Ozaydin, Rumeysa and Mazein, Alexander and Auffray, Charles and Babur, {\"O}zg{\"u}n and Demir, Emek and Dogrusoz, Ugur},
  journal={Bioinformatics},
  volume={37},
  number={10},
  pages={1475--1477},
  year={2021},
  doi={10.1093/bioinformatics/btaa850}
}

@article{xu2023sbmldiagrams,
  title={{SBMLDiagrams}: a python package to process and visualize {SBML} layout and render},
  author={Xu, Jin and Jiang, Jessie and Sauro, Herbert M},
  journal={Bioinformatics},
  volume={39},
  number={1},
  pages={btac730},
  year={2023},
  doi={10.1093/bioinformatics/btac730}
}

@article{czauderna2010sbgned,
  title={Editing, validating and translating of {SBGN} maps},
  author={Czauderna, Tobias and Klukas, Christian and Schreiber, Falk},
  journal={Bioinformatics},
  volume={26},
  number={18},
  pages={2340--2341},
  year={2010},
  doi={10.1093/bioinformatics/btq407}
}

@article{gawron2016minerva,
  title={{MINERVA}---a platform for visualization and curation of molecular interaction networks},
  author={Gawron, Piotr and Ostaszewski, Marek and Satagopam, Venkata and Gebel, Stephan and Mazein, Alexander and Kuzma, Michal and Zorzan, Simone and McGee, Fintan and Otjacques, Beno{\^\i}t and Balling, Rudi and Schneider, Reinhard},
  journal={npj Systems Biology and Applications},
  volume={2},
  pages={16020},
  year={2016},
  doi={10.1038/npjsba.2016.20}
}

@misc{wu2026metacarto,
  title={{MetaCarto}: biologically faithful automatic layout for genome-scale metabolic maps},
  author={Wu, Tianyu},
  howpublished={openRxiv preprint},
  year={2026},
  doi={10.64898/2026.09.19.752882}
}

@article{brunk2018recon3d,
  title={{Recon3D} enables a three-dimensional view of gene variation in human metabolism},
  author={Brunk, Elizabeth and Sahoo, Swagatika and Zielinski, Daniel C and Altunkaya, Ali and Dr{\"a}ger, Andreas and Mih, Nathan and Gatto, Francesco and Nilsson, Avlant and Preciat Gonzalez, German Andres and Aurich, Maike Kathrin and Prli{\'c}, Andreas and Sastry, Anand and Danielsdottir, Anna D and Heinken, Almut and Noronha, Alberto and Rose, Peter W and Burley, Stephen K and Fleming, Ronan M T and Nielsen, Jens and Thiele, Ines and Palsson, Bernhard O},
  journal={Nature Biotechnology},
  volume={36},
  number={3},
  pages={272--281},
  year={2018},
  doi={10.1038/nbt.4072}
}

@article{olivier2018fbc,
  title={{SBML} Level 3 Package: Flux Balance Constraints version 2},
  author={Olivier, Brett G. and Bergmann, Frank T.},
  journal={Journal of Integrative Bioinformatics},
  volume={15},
  number={1},
  pages={20170082},
  year={2018},
  doi={10.1515/jib-2017-0082}
}

@article{uniprot2025,
  title={{UniProt}: the Universal Protein Knowledgebase in 2025},
  author={{The UniProt Consortium}},
  journal={Nucleic Acids Research},
  volume={53},
  number={D1},
  pages={D609--D617},
  year={2025},
  doi={10.1093/nar/gkae1010}
}

@misc{henderson2016conservation,
  title={Conservation analysis of genome-scale biochemical networks},
  author={Henderson, Nick W and Ma, Ding and Sun, Yuekai and Saunders, Michael and Sun, Yuekai and Fleming, Ronan M.T. and  Thiele, Ines},
  howpublished = {\url{https://web.stanford.edu/group/SOL/multiscale/papers/conservation06.pdf}},
  year={2016}
}

@article{king2016bigg,
  title={{BiGG} Models: A platform for integrating, standardizing and sharing genome-scale models},
  author={King, Zachary A. and Lu, Justin and Dr{\"a}ger, Andreas and Miller, Philip and Federowicz, Stephen and Lerman, Joshua A. and Ebrahim, Ali and Palsson, Bernhard O. and Lewis, Nathan E.},
  journal={Nucleic Acids Research},
  volume={44},
  number={D1},
  pages={D515--D522},
  year={2016},
  doi={10.1093/nar/gkv1049}
}

@article{vallabhajosyula2006conservation,
  title={Conservation analysis of large biochemical networks},
  author={Vallabhajosyula, Ravishankar Rao and Chickarmane, Vijay and Sauro, Herbert M.},
  journal={Bioinformatics},
  volume={22},
  number={3},
  pages={346--353},
  year={2006},
  doi={10.1093/bioinformatics/bti800}
}

@article{reder1988metabolic,
  title={Metabolic control theory: a structural approach},
  author={Reder, Christine},
  journal={Journal of Theoretical Biology},
  volume={135},
  number={2},
  pages={175--201},
  year={1988},
  doi={10.1016/S0022-5193(88)80073-0}
}

@book{herbert2019systems,
  title={Systems biology: an introduction to metabolic control analysis},
  author={HERBERT, M SAURO},
  year={2019},
  publisher={Ambrosius Publishing}
}

@article{fell1985metabolic,
  title={Metabolic control and its analysis. {A}dditional relationships between elasticities and control coefficients},
  author={Fell, David A. and Sauro, Herbert M.},
  journal={European Journal of Biochemistry},
  volume={148},
  number={3},
  pages={555--561},
  year={1985},
  doi={10.1111/j.1432-1033.1985.tb08876.x}
}

@article{klamt2007structural,
  title={Structural and functional analysis of cellular networks with {CellNetAnalyzer}},
  author={Klamt, Steffen and Saez-Rodriguez, Julio and Gilles, Ernst D.},
  journal={BMC Systems Biology},
  volume={1},
  pages={2},
  year={2007},
  doi={10.1186/1752-0509-1-2}
}

@article{novere2005miriam,
  title={Minimum information requested in the annotation of biochemical models ({MIRIAM})},
  author={Le Nov{\`e}re, Nicolas and Finney, Andrew and Hucka, Michael and Bhalla, Upinder S. and Campagne, Fabien and Collado-Vides, Julio and Crampin, Edmund J. and Halstead, Matt and Klipp, Edda and Mendes, Pedro and others},
  journal={Nature Biotechnology},
  volume={23},
  number={12},
  pages={1509--1515},
  year={2005},
  doi={10.1038/nbt1156}
}

@article{juty2012identifiers,
  title={Identifiers.org and {MIRIAM} {Registry}: community resources to provide persistent identification},
  author={Juty, Nick and Le Nov{\`e}re, Nicolas and Laibe, Camille},
  journal={Nucleic Acids Research},
  volume={40},
  number={D1},
  pages={D580--D586},
  year={2012},
  doi={10.1093/nar/gkr1097}
}

@article{hastings2016chebi,
  title={{ChEBI} in 2016: Improved services and an expanding collection of metabolites},
  author={Hastings, Janna and Owen, Gareth and Dekker, Adriano and Ennis, Marcus and Kale, Namrata and Muthukrishnan, Venkatesh and Turner, Steve and Swainston, Neil and Mendes, Pedro and Steinbeck, Christoph},
  journal={Nucleic Acids Research},
  volume={44},
  number={D1},
  pages={D1214--D1219},
  year={2016},
  doi={10.1093/nar/gkv1031}
}

@article{bansal2022rhea,
  title={Rhea, the reaction knowledgebase in 2022},
  author={Bansal, Parit and Morgat, Anne and Axelsen, Kristian B. and Muthukrishnan, Venkatesh and Coudert, Elisabeth and Aimo, Lucila and Hyka-Nouspikel, Nevila and Gasteiger, Elisabeth and Kerhornou, Arnaud and Neto, Teresa Batista and others},
  journal={Nucleic Acids Research},
  volume={50},
  number={D1},
  pages={D693--D700},
  year={2022},
  doi={10.1093/nar/gkab1016}
}

@inproceedings{jupp2015ols,
  author    = {Jupp, Simon and Burdett, Tony and Malone, James and
               Leroy, Catherine and Pearce, Matt and McMurry, Julie and
               Parkinson, Helen},
  title     = {A new Ontology Lookup Service at {EMBL-EBI}},
  booktitle = {Proceedings of the 8th International Conference on
               Semantic Web Applications and Tools for Life Sciences
               ({SWAT4LS})},
  series    = {CEUR Workshop Proceedings},
  volume    = {1546},
  pages     = {118--119},
  year      = {2015}
}

@article{kim2023pubchem,
  title={{PubChem} 2023 update},
  author={Kim, Sunghwan and Chen, Jie and Cheng, Tiejun and Gindulyte, Asta and He, Jia and He, Siqian and Li, Qingliang and Shoemaker, Benjamin A. and Thiessen, Paul A. and Yu, Bo and Zaslavsky, Leonid and Zhang, Jian and Bolton, Evan E.},
  journal={Nucleic Acids Research},
  volume={51},
  number={D1},
  pages={D1373--D1380},
  year={2023},
  doi={10.1093/nar/gkac956}
}

@article{moretti2021metanetx,
  title={{MetaNetX/MNXref}: unified namespace for metabolites and biochemical reactions in the context of metabolic models},
  author={Moretti, S{\'e}bastien and Tran, Van Du T. and Mehl, Florence and Ibberson, Mark and Pagni, Marco},
  journal={Nucleic Acids Research},
  volume={49},
  number={D1},
  pages={D570--D574},
  year={2021},
  doi={10.1093/nar/gkaa992}
}

\end{document}